\documentclass[a4paper,12pt]{article}
\usepackage{latexsym}
\usepackage[dvips]{graphicx}
\usepackage{amssymb,amsmath}
\begin{document}

\title{The Membrane Model of Black Holes and Applications}
\author{Norbert Straumann}

\maketitle

\begin{abstract}
In my lectures I shall give an introduction to the "membrane model" of black 
holes (BHs). This is not a new theory, but a useful reformulation of the 
standard relativistic theory of BH --as far as physics outside the horizon is 
concerned-- which is much closer to the intuition we have gained from other 
fields of physics. While the basic equations look less elegant, it has the 
advantage that we can understand astrophysical processes near a BH much 
more easily. In the membrane model one first splits the elegant 4-dimensional
physical laws into space and time (3+1 splitting). For a stationary BH 
there is a preferred decomposition. Relative to this the dynamical variables 
(electromagnetic fields, etc.) become quantities of an "absolute space" which 
evolve as functions of an "absolute time", as we are accustomed to from 
nonrelativistics physics. We shall see, for example, that the 3+1 splitting 
brings Maxwell's equations into a form which resembles the familiar form of 
Maxwell's equations for moving conductors. We can then use the pictures 
and the experience from ordinary electrodynamics.

In a second step, one replaces the boundary conditions at the horizon by 
physical properties (electric conductivity, etc.) of a fictitious membrane. This 
procedure is completely adequate as long as one is not interested in fine 
details very close to the horizon. The details of this boundary layer are, 
however, completely irrelevant for astrophysical applications.

The following points will be discussed in detail:
\begin{itemize}
\item Solutions of Maxwell's equations in a Kerr background
\item Space-time splittings
\item The horizon as a conducting membrane
\item Magnetic energy extraction from a BH
\item BH's as current generators or rotators of electric motors
\item The Blandford-Znajek process
\item Stationary axisymmetric electrodynamics for force-free fields
\end{itemize}
\end{abstract}

\section{Introduction}
In these lectures I give an introduction to what is called \emph{the 
membrane model of black holes} (BHs). This is not a new theory, but a 
convenient reformulation of the standard relativistic theory of BHs -- 
as far as physics \emph{outside} the horizon is concerned --, which is 
much closer to the intuition we have gained from other fields of 
physics (see \cite{1}, for a general reference). While the basic 
equations look less elegant, it has the advantage that we can 
understand astrophysical processes near a BH much more easily. For an 
analogy, imagine you would have to explain how a Tokomak works by 
using the language and pictures of special relativity (SR), i.e., by using 
the electromagnetic field tensor and 4-dimensional pictures of plasma 
flows. I would not know how to do this and how to get, for instance, 
an understanding of even the simplest plasma instabilities. A closer 
analogy would be to translate relevant studies of the electrodynamics 
of pulsars into a 4-dimensional language. The basic equations look 
beautiful, but it would be hard to understand anything. (You may say 
that radio pulsars are anyhow not understood.)

In the membrane model (often called \emph{membrane paradigm} \cite{1}) 
one first splits the elegant 4-dimensional physical laws of general 
relativity (GR) into space and time (3+1 splitting). For a general 
situation this can be done in many ways (reflecting the gauge freedom 
in GR) since there is no canonical fibration of spacetime by level 
surfaces of constant time. However, for a stationary BH there is a 
preferred decomposition. Relative to this the dynamical variables 
(electromagnetic fields, etc) become quantities on an \emph{absolute 
space} which evolve as functions of an \emph{absolute time}, as we 
are accustomed to from nonrelativistic physics. We shall see, for 
example, that the 3+1 splitting brings Maxwell's equations into a 
form which resembles the familiar form of Maxwell's equations for 
moving conductors. We can then use the pictures and the experience from 
ordinary electrodynamics.

In a second step one replaces the boundary conditions at the horizon 
by physical properties (electric conductivity, etc) of a 
\emph{fictitious membrane}. This procedure is completely adequate as 
long as one is not interested in fine details \emph{very close} to the 
horizon. The details of this boundary layer are, however, completely 
irrelevant for astrophysical applications. (The situation is similar 
to many problems in electrodynamics, where one replaces the real 
surface properties of a conductor and other media by idealized 
boundary conditions.)

The program of these lectures  is as follows. First I will discuss 
the 3+1 splitting of the spacetime of a stationary rotating BH and of 
Maxwell's equations outside its horizon. We shall see that this can be 
achieved very smoothly by using the calculus of differential forms. As 
an illustration and for later use we shall apply these tools for a 
discussion of an exact solution of Maxwell's equation on a Kerr 
background, which describes an asymptotically homogeneous magnetic 
field. We shall then derive the electromagnetic properties of the 
fictitious membrane that simulate the boundary conditions at the 
horizon. Here, I can offer a much simpler derivation than has been 
given so far in the literature. As an important example of a physical 
process relatively close to a BH I will treat in detail the magnetic 
energy extraction of a hole's rotational energy. Blandford and Znajek 
have first pointed out the possible relevance of this mechanism for an 
understanding of active galactic nuclei. It may well play an important 
role in the formation of energetic jets. The Blandford-Znajek process 
could also be important for explaining gamma-ray-bursts, because it 
may energize a Poynting-dominated outflow.

\begin{figure}[htbp]
        \centerline{\includegraphics[width=10cm, height=10cm]{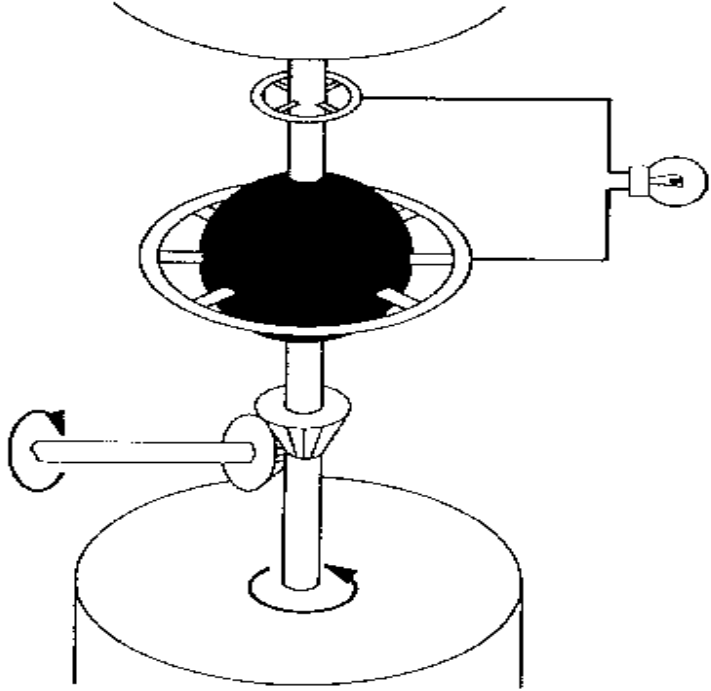}}
        \caption{Electric generator whose physics is similar to the 
        electrodynamics of black holes in external magnetic fields.}
        \protect\label{strau:Fig.1.1}
\end{figure}

I hope to show you that the physics involved is not very different 
from that behind the electric generator in Fig. \ref{strau:Fig.1.1}.

\section{Space-Time Splitting of Electrodynamics}
I describe now the 3+1 splitting of the general relativistic 
Maxwell equations on a stationary spacetime $(M,^{(4)}\!\boldsymbol{g})$. 
Most of what follows could easily be generalized to spacetimes which 
admit a foliation by spacelike hypersurfaces (see, e.g., Ref. \cite{2}), 
but this is not needed in what follows.

Slightly more specifically, we shall assume that globally $M$ is a product 
$\boldsymbol{R} \times \Sigma$, such that the natural coordinate $t$ 
of $\boldsymbol{R}$ is adapted to the Killing field $k$, i.e., 
$k=\partial_{t}$. We decompose the Killing field into normal and 
parallel components relative to the ``absolute space'' 
$(\Sigma,\boldsymbol{g})$, $\boldsymbol{g}$ being the induced metric 
on $\Sigma$,
\begin{equation}
        \partial_{t} = \alpha\,u + \beta.
        \label{strau:1}
\end{equation}
Here $u$ is the unit normal field and $\beta$ is tangent to $\Sigma$. 
This is what one calls the decomposition into lapse and shift; 
$\alpha$ is the \emph{lapse function} und $\beta$ the \emph{shift 
vector field}. We shall usually work with adapted coordinates 
$(x^{\mu})=(t,x^{i})$, where $\{x^{i}\}$ is a coordinate system on 
$\Sigma$. Let $\beta = \beta^{i}\partial_{i} \quad 
(\partial_{i}=\partial/\partial x^{i})$, and consider the basis of 
1-forms
\begin{equation}
        \alpha\,dt, \qquad dx^{i} + \beta^{i}dt.
        \label{strau:2}
\end{equation}
One verifies immediately, that this is dual to the basis 
$\{u,\partial_{i}\}$ of vector fields. Since $u$ is perpendicular to 
the tangent vectors $\partial_{i}$ of $\Sigma$, the 4-metric has the 
form
\begin{equation}
        ^{(4)}\boldsymbol{g} = -\alpha^{2}dt^{2} + 
        g_{ij}\left(dx^{i}+\beta^{i}dt\right)
        \left(dx^{j}+\beta^{j}dt\right),
        \label{strau:3}
\end{equation}
where $g_{ij}dx^{i}dx^{j}$ is the induced metric $\boldsymbol{g}$ on 
$\Sigma$. Clearly, $\alpha$, $\beta$, and $\boldsymbol{g}$ are all 
time-independent quantities on $\Sigma$.

For what follows, I would like to change this setup slightly by 
using, instead of $\partial_{i}$ and $dx^{i}$, a dual orthonormal 
pair $\{e_{i}\}$ and $\{\vartheta^{i}\}$ on $\Sigma$. Instead of 
(\ref{strau:2}), we have then the \emph{orthonormal} tetrad
\begin{equation}
        \theta^{0} = \alpha dt,\qquad \theta^{i} = \vartheta^{i}+\beta^{i}dt,
        \label{strau:4}
\end{equation}
where now $\beta=\beta^{i}e_{i}$. This is dual to the orthonormal frame
\begin{equation}
        e_{0} = u = \frac{1}{\alpha}(\partial_{t}-\beta), \quad e_{i}.
        \label{strau:5}
\end{equation}
The tetrad $\{\theta^{\mu}\}$ describes the reference frames of 
so-called FIDOs, for \emph{fiducial observers}. Their 4-velocity is 
thus perpendicular to the absolute space $\Sigma$.

Relative to these observers we have for the electromagnetic field 
tensor $F$ (2-form) the same decomposition as in SR:
\begin{equation}
        F = E \wedge \theta^{0} + B,
        \label{strau:6}
\end{equation}
where $E$ is the electric 1-form $E=E_{i}\theta^{i}$ and $B$ the 
magnetic 2-form $B=\frac{1}{2}B_{ij}\,\theta^{i} \wedge \theta^{j}$. 
($E_{i}$, $B_{ij}$ are the field strengths measured by the FIDOs.)

In a second step  we decompose $E$ and $B$ relative to absolute space 
and absolute time. We have, using (\ref{strau:5}),
\begin{equation}
                E = E_{i}\theta^{i} = E_{i}\left(\vartheta^{i}+\beta^{i}dt\right) = \mathcal{E} + 
                i_{\beta}\,\mathcal{E}\,dt,
        \label{strau:7}
\end{equation}
where
\begin{equation}
        \mathcal{E} = E_{i}\vartheta^{i}, \qquad i_{\bullet}:\quad 
        \textrm{interior product.}
        \label{strau:8}
\end{equation}
Similarly,
\begin{equation}
        B = \mathcal{B}+dt\wedge i_{\beta}\,\mathcal{B}, \quad 
                \mathcal{B}=\frac{1}{2}B_{ij}\,\vartheta^{i}\wedge \vartheta^{j}.
        \label{strau:9}
\end{equation}
Together we arrive at the following 3+1 decomposition of $F$:
\begin{equation}
                F=\mathcal{B}+(\alpha \,\mathcal{E}-i_{\beta}\,\mathcal{B})\wedge dt.
        \label{strau:10}
\end{equation}

>From this the 3+1 splitting of the homogeneous Maxwell equations is 
readily obtained: $dF=0$ gives
\begin{displaymath}
        \mathbf{d}\mathcal{B}+dt\wedge \partial_{t}\,\mathcal{B} 
        +\mathbf{d}(\alpha\,\mathcal{E})\wedge dt - 
        \mathbf{d}(i_{\beta}\,\mathcal{B})\wedge dt =0.
\end{displaymath}
Here $\mathbf{d}$ denotes the differential on $\Sigma$. This gives 
the two equations
\begin{displaymath}
        \mathbf{d}\mathcal{B}=0,\qquad \mathbf{d}(\alpha\,\mathcal{E})+ 
        \partial_{t}\,\mathcal{B} = \mathbf{d}(i_{\beta}\,\mathcal{B})
\end{displaymath}
or, with the Cartan identity $L_{\beta}=\mathbf{d}\circ 
i_{\beta} + i_{\beta}\circ \mathbf{d}$,
\begin{equation}
        \mathbf{d}\mathcal{B}=0,\qquad \mathbf{d}(\alpha\,\mathcal{E}) + 
        (\partial_{t} - L_{\beta})\mathcal{B} =0.
        \label{strau:11}
\end{equation}
The second equation describes Faraday's induction law in a 
gravitational field. It will be of crucial importance in later 
sections. Note, in particular, the coupling of the 
$\mathcal{B}$-field to the shift through the Lie derivative.

Let us also decompose the representation of $F$ by a potential, $F=dA$. 
We have, using again (\ref{strau:4}),
\begin{eqnarray*}
        A & =& A_{\mu}\,\theta^{\mu} = \alpha A_{0}\,dt + 
        A_{i}(\vartheta^{i} + \beta^{i}\,dt)  \\
         & =& (\alpha A_{0}+i_{\beta}\,\mathcal{A})dt + \mathcal{A}, \qquad 
        \mathcal{A}=A_{i}\,\vartheta^{i}.
\end{eqnarray*}
Thus
\begin{equation}
                A=-\phi\,dt + \mathcal{A},
        \label{strau:12}
\end{equation}
where
\begin{equation}
        \phi=-(\alpha A_{0}+i_{\beta}\,\mathcal{A}).
        \label{strau:13}
\end{equation}
This gives
\begin{displaymath}
        dA=-\mathbf{d}\phi \wedge dt + \mathbf{d}\mathcal{A} + dt \wedge 
        \partial_{t}\,\mathcal{A},
\end{displaymath}
which is of the form (\ref{strau:10}), with
\begin{equation}
        \mathcal{B}=\mathbf{d}\mathcal{A}, \qquad 
        \alpha\,\mathcal{E}=-\mathbf{d}\phi-\partial_{t}\,\mathcal{A}+ 
        i_{\beta}\,\mathbf{d}\mathcal{A}.
        \label{strau:14}
\end{equation}
Apart from the last term, this is what one is used to.

Now, we turn to the inhomogeneous Maxwell equation
\begin{equation}
        d*F=4\pi \mathcal{S}.
        \label{strau:15}
\end{equation}
We need first the Hodge-dual of (\ref{strau:10}). We decompose $*F$ 
similarly to (\ref{strau:6}):
\begin{equation}
        *F=-H \wedge \theta^{0} + D,
        \label{strau:16}
\end{equation}
which can be viewed as a definition of $H$ and $D$. Comparison with 
(\ref{strau:6}) shows, that
\begin{eqnarray}
    H & = & B_{i}\,\theta^{i}, \qquad B_{1}= B_{23},\:\textrm{etc,} 
        \nonumber \\
        D & = & E_{1}\,\theta^{2} \wedge \theta^{3} + 
        E_{2}\,\theta^{3}\wedge \theta^{1} + E_{3}\,\theta^{1}\wedge 
        \theta^{2}.
        \label{strau:17}
\end{eqnarray}
With (\ref{strau:4}) we find ($\boldsymbol{*}$ denotes the Hodge-dual on 
$\Sigma$)
\begin{eqnarray}
        H & = & \mathcal{H} + i_{\beta}\,\mathcal{H}\wedge dt, \qquad 
        \mathcal{H} = \boldsymbol{*}\mathcal{B}, \nonumber \\
        D & = & \mathcal{D} - i_{\beta}\,\mathcal{D} \wedge dt, \qquad 
        \mathcal{D} = \boldsymbol{*}\mathcal{E}.
        \label{strau:18}
\end{eqnarray}
If this is inserted into (\ref{strau:16}), we obtain
\begin{equation}
        *F=\mathcal{D} - (\alpha \mathcal{H} + i_{\beta}\,\mathcal{D})\wedge 
        dt.
        \label{strau:19}
\end{equation}

The dual $J=*\mathcal{S}$ of the current 3-form can be decomposed as 
in SR
\begin{equation}
        J=\rho_{el}\,\theta^{0} + j_{k}\,\theta^{k},
        \label{strau:20}
\end{equation}
where $\rho_{el}$ is the electric charge density and $j^{k}$ is the 
electric current density relative to the FIDOs. We use them to 
introduce the following quantities on the absolute space
\begin{equation}
        \rho = \rho_{el}\,\vartheta^{1}\wedge \vartheta^{2} \wedge 
        \vartheta^{3}, \qquad j=j_{k}\,\vartheta^{k}, \qquad \mathcal{J}=\boldsymbol{*}j.
        \label{strau:21}
\end{equation}

Using the notation $\eta^{\mu}:=*\theta^{\mu}$, we can decompose 
$\mathcal{S}$ as follows
\begin{eqnarray*}
        \mathcal{S} & = & \rho_{el}\,\eta^{0} + j_{k}\,\eta^{k}  \\
         & = & \rho_{el}\,\theta^{1}\wedge \theta^{2} \wedge \theta^{3} - 
         (j_{1}\,\theta^{2} \wedge \theta^{3}+\ldots)\wedge \theta^{0} \\
         & = & \rho + \rho_{el}(\beta^{1}\,\vartheta^{2}\wedge 
         \vartheta^{3}+\ldots) \wedge dt - \alpha \,(j_{1}\,\vartheta^{2}\wedge 
         \vartheta^{3}+\ldots)\wedge dt.
\end{eqnarray*}
Thus
\begin{equation}
        \mathcal{S} = \rho + (i_{\beta}\,\rho - \alpha \, \mathcal{J})\wedge dt.
        \label{strau:22}
\end{equation}
Inserting this and (\ref{strau:19}) into (\ref{strau:15}) leads to
\begin{eqnarray*}
        d*F & = & \mathbf{d}\mathcal{D}+dt \wedge \partial_{t}\,\mathcal{D} - 
        \mathbf{d}(\alpha \mathcal{H}) \wedge dt - 
        \mathbf{d}(i_{\beta}\,\mathcal{D}) \wedge dt  \\
         & = & 4\pi \rho + 4\pi (i_{\beta}\,\rho - \alpha\, 
         \mathcal{J})\wedge dt,
\end{eqnarray*}
and hence to the following 3+1 split of the inhomogeneous Maxwell 
equation
\begin{equation}
        \mathbf{d}\mathcal{D} = 4\pi \rho, \qquad \mathbf{d}(\alpha 
        \mathcal{H}) = (\partial_{t}-L_{\beta})\mathcal{D} + 4 \pi \alpha\,\mathcal{J}.
        \label{strau:23}
\end{equation}

>From these laws one obtains immediately the local conservation law of 
the electric charge (use that $\mathbf{d}$ commutes with $L_{\beta}$):
\begin{equation}
        (\partial_{t}-L_{\beta})\rho + \mathbf{d}(\alpha\,\mathcal{J}) = 0.
        \label{strau:24}
\end{equation}
This follows, of course, also from $d\mathcal{S}=0$ and the 
decomposition (\ref{strau:22}).

\begin{center}
        \large{\textbf{Integral Formulas}}
\end{center}

As is well-known from ordinary electrodynamics, it is often useful to 
write the basic laws (\ref{strau:11}), (\ref{strau:23}), and (\ref{strau:24}) in 
integral forms. Consider, for instance, the induction law in 
(\ref{strau:11}). If we integrate this over a surface area $\mathcal{A}$, 
which is \emph{at rest} relative to the absolute space, we obtain with 
Stokes' theorem $(\mathcal{C}:= \partial\mathcal{A})$
\begin{displaymath}
        \oint_{\mathcal{C}}^{}\alpha\,\mathcal{E} = 
        -\frac{d}{dt}\int_{\mathcal{A}}^{}\mathcal{B} + 
        \int_{\mathcal{A}}^{}L_{\beta}\,\mathcal{B}. 
\end{displaymath}
Here, we use 
$L_{\beta}\,\mathcal{B}=\mathbf{d}\,i_{\beta}\,\mathcal{B}$ (since 
$\mathbf{d}\mathcal{B}=0$) and Stokes' theorem once more, with the 
result
\begin{equation}
        \oint_{\mathcal{C}}^{}\alpha\,\mathcal{E} = 
        -\frac{d}{dt}\int_{\mathcal{A}}^{}\mathcal{B} + 
        \oint_{\mathcal{C}}^{}i_{\beta}\,\mathcal{B}. 
        \label{strau:25}
\end{equation}
The left hand side is the electromotive force (EMF) along $\mathcal{C}$. 
The last term is similar to the additional term one encounters in 
Faraday's induction law for moving conductors. It is an expression of 
the coupling of $\mathcal{B}$ to the gravitomagnetic field and plays a 
crucial role in much that follows. This term contributes also for a 
stationary situation, for which (\ref{strau:25}) reduces to
\begin{equation}
        \textrm{EMF}(\mathcal{C}) =\oint_{\mathcal{C}}^{}\alpha\,\mathcal{E} =
         \oint_{\mathcal{C}}i_{\beta}\,\mathcal{B}.
        \label{strau:26}
\end{equation}

The integral form of the Amp\`{e}re-Maxwell law is obtained similarly. 
Integrating the second equation in (\ref{strau:23}), we obtain with the 
Cartan identity $L_{\beta}=\mathbf{d}\circ i_{\beta} + i_{\beta}\circ \mathbf{d}$
and Gauss' law (first equation in (\ref{strau:23})):
\begin{equation}
        \oint_{\mathcal{C}}^{}(\alpha\,\mathcal{H}+i_{\beta}\,\mathcal{D}) = \frac{d}{dt} 
        \int_{\mathcal{A}}^{}\mathcal{D} +4\pi 
        \int_{\mathcal{A}}^{}(\alpha\,\mathcal{J}-i_{\beta}\,\rho ).
        \label{strau:27}
\end{equation}

The integral form of charge conservation is obtained by integrating 
(\ref{strau:24}) over a volume $\mathcal{V}$ which is at rest relative to 
absolute space:
\begin{equation}
        \frac{d}{dt}\int_{\mathcal{V}}^{}\rho = -\int_{\partial 
        \mathcal{V}}^{}(\alpha\,\mathcal{J}-i_{\beta}\,\rho)
        \label{strau:28}
\end{equation}
(note that $L_{\beta}\,\rho=\mathbf{d}\,i_{\beta}\,\rho)$.

One could, of course, also derive integral formulas for moving volumes 
and surface areas (exercise).

\begin{center}
        \large{\textbf{Vector Analytic Formulation}}
\end{center}

The similarity of the basic laws in the 3+1 split with ordinary 
electrodynamics becomes even closer if we write everything in vector 
analytic form. I give a dictionary between the two formulations that is 
valid for any 3-dimensional Riemannian manifold 
$(\Sigma,\boldsymbol{g})$.

The metric $\boldsymbol{g}$ defines natural isomorphisms $\sharp$ 
and $\flat$ between the sets of 1-forms, $\Lambda^{1}(\Sigma)$, and 
vector fields, $\mathcal{X}(\Sigma)$. In addition, the volume form 
$\eta$, belonging to the metric, defines an isomorphism between $\mathcal{X}(\Sigma)$
and the space of 2-forms, $\Lambda^{2}(\Sigma)$, given by
\begin{equation}
        \vec{B}\longmapsto\mathcal{B}=i_{\vec{B}}\,\eta.
        \label{strau:29}
\end{equation}
We have the following commutative diagram, in which $\boldsymbol{*}$ denotes,
as always, the Hodge-dual:
\begin{displaymath}
        \begin{picture}(8,3) 
\put(0,2.5){$\Lambda^{1}(\Sigma)$}   \put(2.2,0.5){\vector(-2,3){1.1}}
\put(4,2.5){$\Lambda^{2}(\Sigma)$}   \put(0.8,2.1){\vector(2,-3){1.1}}
\put(2,0){$\mathcal{X}(\Sigma)$}    
\put(4,1.2){$ i_{\bullet}\,\eta$}
%\put(5.5,1.25){($\vec{B}\longmapsto\mathcal{B}=i_{\vec{B}}\,\eta.)$}
\put(3.5,2.5){$\vector(-1,0){2}$}   \put(3,0.5){\vector(2,3){1.1}}
\put(1.5,2.7){$\vector(1,0){2}$}
\put(1,1){$ \sharp $}  \put(1.9,1.5){$ \flat$}
\put(2.5,2.75){$\boldsymbol{*}$} \put(2.5,2.1){$\boldsymbol{*}$}
\end{picture}
\end{displaymath}
>From this one can read off, for instance,
\begin{equation}
        i_{\vec{v}}\,\eta = \boldsymbol{*} v \qquad (\vec{v}\in\mathcal{X}(\Sigma),\; 
        v:\:(\vec{v})^{\flat}). 
        \label{strau:30}
\end{equation}
The cross product and the wedge product are related as follows:
\begin{displaymath}
        \begin{array}{ccc}
                \Lambda^{1}(\Sigma)\times \Lambda^{1}(\Sigma)& 
                \stackrel{\wedge}{\longrightarrow} & \Lambda^{2}(\Sigma)  \\
                \updownarrow &  & \updownarrow  \\ \mathcal{X}(\Sigma) \times 
        \mathcal{X}(\Sigma) &\stackrel{\times}{\longrightarrow} & \mathcal{X}(\Sigma)  
   \end{array}
\end{displaymath}
In particular, we have
\begin{equation}
        i_{\vec{v}\times \vec{w}}\,\eta = v\wedge w.
        \label{strau:31}
\end{equation}
With the help of the next commutative diagram one can reduce many of 
the vector analytic identities to $d \circ d =0$.
\begin{displaymath}
        \begin{array}{ccccccccccc}
                0 & \longrightarrow & \Lambda^{0}(\Sigma) & \stackrel{\textbf{d}}{\longrightarrow} 
                &\Lambda^{1}(\Sigma)  &\stackrel{\textbf{d}}{\longrightarrow}  & \Lambda^{2}(\Sigma)
                 &\stackrel{\textbf{d}}{\longrightarrow} & \Lambda^{3}(\Sigma) & \longrightarrow & 
                 0  \\
                 &  & \Arrowvert &  & {\scriptstyle \flat} \uparrow \downarrow {\scriptstyle \sharp}
                  & & \uparrow 
                 {\scriptstyle i_{\bullet}\,\eta} &  &\uparrow {\scriptstyle \bullet \,\eta}  &  &   \\
                0 & \longrightarrow &\Lambda^{0}(\Sigma)  & \stackrel{\textrm{grad}}{\longrightarrow}
                 & \mathcal{X}(\Sigma) & \stackrel{\textrm{curl}}{\longrightarrow} & \mathcal{X}(\Sigma) &
                  \stackrel{\textrm{div}}{\longrightarrow} & \Lambda^{0}(\Sigma) & \longrightarrow & 0.
        \end{array}
\end{displaymath}
We can read off, for example,
\begin{equation}
                i_{\textrm{curl}\,\vec{v}}\,\eta = \mathbf{d}v.
        \label{strau:32}
\end{equation}
For a 1-form $w$ with vector field $\vec{w}=w^{\sharp}$,we have the 
translation
\begin{equation}
        i_{\vec{v}}\,\mathbf{d}w \stackrel{(\ref{strau:32})}{=} i_{\textrm{curl}\,
        \vec{v}}\,i_{\textrm{curl}\,\vec{w}}\,\eta = 
        [(\textrm{curl}\,\vec{w})\times \vec{v}]^{\flat}.
        \label{strau:33}
\end{equation}
Here, we made use of the algebraic relation
\begin{equation}
        i_{\vec{v}}\,i_{\vec{u}}\,\eta \stackrel{(\ref{strau:30})}{=} 
        i_{\vec{v}}\,\boldsymbol{*}u = \boldsymbol{*}(u\wedge v) \stackrel{(\ref{strau:31})}{=} 
        (\vec{u}\times \vec{v})^{\flat}.
        \label{strau:34}
\end{equation}

We need also the translation of the Lie derivative of a 1-form $w$:
\begin{displaymath}
        L_{\vec{v}}\,w = \mathbf{d} \underbrace{i_{\vec{v}}\,w}_{(\vec{v},\vec{w})} 
         + i_{\vec{v}}\,\mathbf{d}w \stackrel{(\ref{strau:33})}{=} 
        \mathbf{d}(\vec{v},\vec{w}) + [(\textrm{curl}\,\vec{w})\times \vec{v}]^{\flat}.
\end{displaymath}
Thus,
\begin{equation}
        L_{\vec{v}}\,w = \{\textrm{grad}\,(\vec{v} ,\vec{w}) + (\textrm{curl}\,\vec{w})\times 
        \vec{v}\}^{\flat}.
        \label{strau:35}
\end{equation}

Similarly, we have for a 2-form $\mathcal{B}=i_{\vec{B}}\,\eta$:
\begin{eqnarray*}
    L_{\vec{v}}\,\mathcal{B}& =& L_{\vec{v}}\,i_{\vec{B}}\,\eta = 
        i_{\vec{v}}\,\underbrace{\mathbf{d}\,i_{\vec{B}}\,\eta}_{\textrm{div}\,\vec{B}\,\eta} 
        + \;\mathbf{d}\underbrace{i_{\vec{v}}\,i_{\vec{B}}\,\eta}_{(\vec{B}\times 
        \vec{v})^{\flat}}  \\
        &\stackrel{(\ref{strau:32})}{=}&i_{\{(\textrm{div}\,\vec{B})\vec{v}\:+\:\textrm{curl}\,(\vec{B}\times 
        \vec{v})\}}\,\eta.
\end{eqnarray*}
We have thus the correspondence
\begin{equation}
         L_{\vec{v}}\,\mathcal{B} \longleftrightarrow (\textrm{div}\,\vec{B})\vec{v}\:
         +\:\textrm{curl}\,(\vec{B}\times \vec{v}).
        \label{strau:36}
\end{equation}
Here we have to stress that the right hand side is in general not 
equal to $L_{\vec{v}}\,\vec{B}=[\vec{v},\vec{B}]$ (Lie bracket). This 
comes out as follows:
\begin{eqnarray*}
        L_{\vec{v}}\,\mathcal{B} & = & L_{\vec{v}}\,i_{\vec{B}}\,\eta = 
        \underbrace{[L_{\vec{v}},i_{\vec{B}}]}_{i_{[\vec{v},\vec{B}]}}\,\eta + 
        i_{\vec{B}}\,\underbrace{L_{\vec{v}}\,\eta}_{\textrm{div}\,\vec{v}\,\eta}   \\
         & = & i_{\left\{(\textrm{div}\,\vec{v})\vec{B}\: +\: [\vec{v},\vec{B}] 
         \right\}}\,\eta.
\end{eqnarray*}
The correspondence (\ref{strau:36}) is thus equivalent to
\begin{equation}
        L_{\vec{v}}\,\mathcal{B} \longleftrightarrow L_{\vec{v}}\,\vec{B} + 
        (\textrm{div}\,\vec{v})\,\vec{B}.
        \label{strau:37}
\end{equation}
Only for $\textrm{div}\,\vec{v}=0$ do the Lie derivatives 
$L_{\vec{v}}\,\mathcal{B}$ and $L_{\vec{v}}\,\vec{B}$ correspond to 
each other!

In Maxwell's equations the Lie derivative $L_{\vec{\beta}}\,\mathcal{B}$ 
occurs. This will be replaced in the vector analytic translation by 
$L_{\vec{\beta}}\,\vec{B}$, because $\textrm{div}\,\vec{\beta}$ is (for a stationary 
metric) proportional to the trace of the second fundamental form of 
the time slices and this vanishes for \emph{maximal slicing}. For the 
Kerr solution we are, for instance, in this situation (exercise).

Part of what has been said is summarized for convenience in the table 
below.
\begin{center}
        \underline{Dictionary}
\end{center}
\begin{displaymath}
         \begin{tabular}{|c|c|c|}
                \hline
                calculus of forms & vector analysis & notation  \\
                \hline
                $v \wedge w$ &$ \vec{v} \times \vec{w}$ &$ v=(\vec{v})^{\flat},\:w=(\vec{w})^{\flat}$ \\
                %\hline
                $i_{\vec{v}}\,\mathcal{B},\:i_{\vec{v}}\,\mathcal{D}$ &$ \vec{B} \times 
                \vec{v},\:\vec{E}\times \vec{v}$ & $
                \mathcal{B}=i_{\vec{B}}\,\eta,\:\mathcal{D}=i_{\vec{E}}\,\eta $ \\
                %\hline
                $\mathbf{d}f$ & $\textrm{grad}\,f$& $f:\textrm{ function} $ \\
                %\hline
                $\mathbf{d}v$ & $\textrm{curl}\,\vec{v}$ &   \\
                %\hline
                $\mathbf{d}\mathcal{B},\:\mathbf{d}\mathcal{D}$ &$ \textrm{div}\,\vec{B},
                \:\textrm{div}\,\vec{E}$ &   \\
                %\hline
                $L_{\vec{v}}\,w$ &$ \textrm{grad}\,(\vec{v},\vec{w})-\vec{v}\times      
                \textrm{curl}\,\vec{w}$ & $
                w=(\vec{w})^{\flat}$  \\
                %\hline
                $L_{\vec{v}}\,\mathcal{B},\:    L_{\vec{v}}\,\mathcal{D} $& 
                $(\textrm{div}\,\vec{B})\vec{v} - \textrm{curl}(\vec{v}\times 
                \vec{B}),\:\vec{B}\longleftrightarrow \vec{E}$ &\\
                \hline
    \end{tabular} 
\end{displaymath}

\begin{center}
        \large{\textbf{Summary}}
\end{center}

For reference, we write down once more the 3+1 split of Maxwell's 
equations (\ref{strau:11}) and (\ref{strau:23}) in Cartan's calculus
\begin{eqnarray}
        \mathbf{d}\mathcal{B}=0, &  &\mathbf{d}(\alpha\,\mathcal{E}) + 
        (\partial_{t} - L_{\beta})\mathcal{B} =0, \nonumber \\
        \mathbf{d}\mathcal{D} = 4\pi \rho, &  & \mathbf{d}(\alpha 
        \mathcal{H}) = (\partial_{t}-L_{\beta})\mathcal{D} + 4 \pi \alpha\,\mathcal{J}.
        \label{strau:38}
\end{eqnarray}
The dictionary above allows us to translate these into the vector 
analytic form:
\begin{eqnarray}
        \vec{\nabla}\cdot \vec{B}=0, &  & \vec{\nabla}\times 
        (\alpha\,\vec{E}) + (\partial_{t}-L_{\vec{\beta}})\vec{B}=0, \nonumber \\
        \vec{\nabla}\cdot \vec{E}=4\pi \rho_{el}, &  & \vec{\nabla}\times 
        (\alpha\,\vec{B}) = (\partial_{t}-L_{\vec{\beta}})\vec{E} + 4\pi \alpha \,\vec{j}.
        \label{strau:39}
\end{eqnarray}

\section{Black Hole in a Homogeneous Magnetic Field}
As an instructive example and a useful tool we discuss now an exact 
solution of Maxwell's equations in the Kerr metric, which becomes 
asymptotically a homogeneous magnetic field. This solution can be 
found in a strikingly simple manner \cite{3}.

For any Killing field $K$ one has the following identity
\begin{equation}
        \delta \,d\,K^{\flat}=2\,R(K),
        \label{strau:40}
\end{equation}
where $\delta$ denotes the codifferential and $R(K)$ is the 1-form 
with components $R_{\mu \nu}K^{\nu}$. In components (\ref{strau:40}) is 
equivalent to
\begin{equation}
        K_{\mu \:\:;\alpha}^{\:;\alpha} = -R_{\mu \alpha}\,K^{\alpha}.
        \label{strau:41}
\end{equation}
This form can be obtained by contracting the indices $\sigma$ and 
$\rho$ in the following general equation for a vector field
\begin{displaymath}
        \xi_{\sigma;\rho \mu} - \xi_{\sigma;\mu \rho} = 
        \xi_{\lambda}R^{\lambda}_{\sigma \rho \mu}
\end{displaymath}
and by using the consequence $K^{\sigma}_{;\sigma}=0$ of the Killing 
equation $K_{\sigma;\rho}+K_{\rho;\sigma}=0$.

For a vacuum spacetime we thus have
\begin{equation}
                \delta \,d\,K^{\flat}=0
        \label{strau:42}
\end{equation}
for any Killing field. Hence, the vacuum Maxwell equations are 
satisfied if $F$ is a constant linear combination of the differential 
of Killing fields (their duals, to be precise). For the Kerr metric, 
as for any axially symmetric stationary spacetime, we have two Killing 
fields $k$ and $m$, say; in adapted coordinates these are 
$k=\partial_{t}$ and $m=\partial_{\varphi}$. The Komar formulae 
provide convenient expressions for the total mass $M$ and the total 
angular momentum $J$ of the Kerr BH:
\begin{equation}
        M=-\frac{1}{8\pi}\int_{\infty}*dk^{\flat},\qquad J=\frac{1}{16\pi}
        \int_{\infty}*dm^{\flat}
        \label{strau:43}
\end{equation}
(for $G=1$).

We try the ansatz
\begin{equation}
        F = \frac{1}{2}\,B_{0}\,(dm^{\flat}+2a\,dk^{\flat}) \qquad (B_{0} = 
        \textrm{const}),
        \label{strau:44}
\end{equation}
and choose $a$ such that the total electric charge
\begin{equation}
        Q = -\frac{1}{4\pi}\int_{\infty}*F
        \label{strau:45}
\end{equation}
vanishes. The Komar formulae (\ref{strau:43}) tell us that
\begin{equation}
        Q = -\frac{1}{8\pi}\,B_{0}\,(16\pi J-2a\cdot 8\pi M),
        \label{strau:46}
\end{equation}
and this vanishes if $a=J/M$ (which is the standard meaning of the 
symbol $a$ in the Kerr solution).

Clearly, $F$ is stationary and axisymmetric:
\begin{equation}
        L_{k}\,F = L_{m}\,F = 0,
        \label{strau:47}
\end{equation}
because (dropping $\flat$ from now on)
\begin{displaymath}
        L_{k}\,dk =d\,L_{k}\,k = 0 \qquad (L_{k}\,k = [k,k] = 0),\:\textrm{etc}.
\end{displaymath}

For the further discussion we need the Kerr metric. In 
Boyer-Lindquist coordinates and more or less standard notation it 
has the form (\ref{strau:3}), i.e.,
\begin{equation}
        ^{(4)}\boldsymbol{g} = [-\alpha^{2}\,dt^{2}+g_{\varphi\varphi}(d\varphi + 
        \beta^{\varphi}\,dt)^{2}] + [g_{rr}\,dr^{2} + 
        g_{\vartheta\vartheta}\,d\vartheta^{2}],
        \label{strau:48}
\end{equation}
with only the component $\beta^{\varphi}$ of the shift being $\neq 0$. 
With the abbreviations
\begin{eqnarray}
        \rho^{2}:=r^{2}+a^{2}\cos^{2}\! \vartheta, & \Delta :=  r^{2}-2Mr+a^{2},
         & \Sigma^{2}:=(r^{2}+a^{2})^{2}-a^{2}\Delta\sin^{2}\!\vartheta ,
        \label{strau:49}
\end{eqnarray}
the metric coefficients are
\begin{eqnarray}
        g_{rr}  =  \frac{\rho^{2}}{\Delta}, \quad g_{\vartheta \vartheta}  =  \rho^{2}, & 
         & g_{\varphi \varphi}= \sin^{2}\! \vartheta \, 
         \frac{\Sigma^{2}}{\rho^{2}},
        \nonumber  \\
        g_{tt}  = -1+\frac{2Mr}{\rho^{2}},  &  & g_{t\varphi} = 
        -\frac{2Mra\sin^{2}\!\vartheta}{\rho^{2}},
        \label{strau:50}
\end{eqnarray}
while the lapse and shift are given by
\begin{equation}
        \alpha^{2}=\frac{\rho^{2}}{\Sigma^{2}}\,\Delta, \qquad
        \beta^{\varphi}=-a\,\frac{2Mr}{\Sigma^{2}}.
        \label{strau:51}
\end{equation}

This gives asymptotically
\begin{eqnarray}
        ^{(4)}\boldsymbol{g} &=& 
        -\left[1-\frac{2M}{r}+\mathcal{O}\left(\frac{1}{r^{2}}\right)\right]\,dt^{2} 
        -\left[\frac{4aM}{r}\,\sin^{2}\!\vartheta + 
        \mathcal{O}\left(\frac{1}{r^{2}}\right)\right]\,dt\,d\varphi
        \nonumber  \\
         &  & + \left[1 + \mathcal{O}\left(\frac{1}{r}\right)\right]\,\left[dr^{2} + 
         r^{2}(d\vartheta^{2} + \sin^{2}\!\vartheta\,d\varphi^{2})\right].
        \label{strau:52}
\end{eqnarray}
To establish the connection with our general discussion, we introduce 
here as orthonormal basis of the absolute space naturally
\begin{equation}
        \vartheta^{r}=\sqrt{g_{rr}}\,dr,\quad 
        \vartheta^{\vartheta}=\sqrt{g_{\vartheta\vartheta}}\,d\vartheta, 
        \quad 
        \vartheta^{\varphi}=\sqrt{g_{\varphi\varphi}}\,d\varphi.
        \label{strau:53}
\end{equation}
The shift vector is $\beta = \beta^{\varphi}\,\partial_{\varphi}$.

The angular velocity $\omega$ of the FIDOs, with 4-velocity $u =\frac{1}{\alpha}
(\partial_{t}-\beta^{\varphi}\,\partial_{\varphi})$, is
\begin{equation}
        \omega = \frac{u^{\varphi}}{u^{t}} = -\beta^{\varphi} = 
        -\frac{g_{t\varphi}}{g_{\varphi \varphi}}.
        \label{strau:54}
\end{equation}
The last equality sign implies that the angular momentum of these 
(Bardeen) observers vanishes: $(u,m) = 0$. The FIDOs are, therefore, 
sometimes also called ZAMOs (for \emph{zero angular momentum observers}).

Now, we want to discuss in detail the solution (\ref{strau:44}), which we 
can express in terms of a potential: $F=dA$, with
\begin{equation}
        A = \frac{1}{2}\,B_{0}\,(m+2ak).
        \label{strau:55}
\end{equation}

Let us first look at the asymptotics. The 1-forms 
$k_{\mu}\,dx^{\mu}$, $m_{\mu}\,dx^{\mu}$ belonging to the Killing 
fields ($k_{\mu}=g_{\mu t}$, $m_{\mu}=g_{\mu \varphi}$) are 
asymptotically $k \sim -dt$, $m \sim 
r^{2}\sin^{2}\!\vartheta\,d\varphi$, whence (\ref{strau:44}) gives
\begin{equation}
        F \sim B_{0}\,[\sin \vartheta\, dr \wedge r\sin \vartheta\, d\varphi
        + \cos \vartheta\,rd\vartheta \wedge r\sin \vartheta \,d\varphi].
        \label{strau:56}
\end{equation}
This is a magnetic field in the z-direction whose magnitude is $B_{0}$.

In (\ref{strau:55}) we need
\begin{eqnarray*}
        m + 2ak & = & (g_{\mu\varphi}+2a\,g_{\mu t})\,dx^{\mu} = 
        (g_{t\varphi}+2a\,g_{tt})\,dt + (g_{\varphi\varphi}+2a\,g_{\varphi 
        t})\,d\varphi  \\
         & \stackrel{(\ref{strau:54})}{=} & (-\omega\,g_{\varphi \varphi} 
         +2a\,g_{tt})dt + (g_{\varphi \varphi}-2a\,\omega\,g_{\varphi 
         \varphi})d\varphi.
\end{eqnarray*}
Using the notation \cite{1}
\begin{equation}
        \tilde{\omega}^{2}:=g_{\varphi \varphi},\qquad 
        \tilde{\omega}=\frac{\Sigma}{\rho}\,\sin\!\vartheta
        \label{strau:57}
\end{equation}
we thus have
\begin{displaymath}
        m+2ak = \left[-\omega \tilde{\omega}^{2} + 
        2a\,\left(\omega^{2}\tilde{\omega}^{2}-\alpha^{2}\right)\right]\,dt + 
        \tilde{\omega}^{2}(1-2a\omega)\,d\varphi.
\end{displaymath}
Comparing this with (\ref{strau:12}), we obtain for the potentials
\begin{eqnarray}
        \phi & = & \frac{1}{2}\,B_{0}\,\left[\omega \tilde{\omega}^{2} + 
        2a\,\left(\alpha^{2} -\omega^{2}\tilde{\omega}^{2}\right)\right],
        \label{strau:58}  \\
        \mathcal{A} & = & \mathcal{A}_{\varphi}\,d\varphi, \qquad 
        \mathcal{A}_{\varphi} =  \frac{1}{2}\,B_{0}\,\tilde{\omega}^{2}(1-2a\omega).
        \label{strau:59}
\end{eqnarray}
The fields $\mathcal{E}$ and $\mathcal{B}$ can now be obtained from 
(\ref{strau:14}). We have
\begin{eqnarray}
        \mathcal{B} & = & d\mathcal{A} = \mathcal{A}_{\varphi,r}\,dr\wedge 
        d\varphi + \mathcal{A}_{\varphi,\vartheta}\,d\vartheta \wedge 
        d\varphi  \nonumber  \\
         & = & \frac{1}{\Sigma\,\sin\!\vartheta} 
         \left[\mathcal{A}_{\varphi,\vartheta}\,\underbrace{\vartheta^{2}\wedge 
         \vartheta^{3}}_{\boldsymbol{*}\vartheta^{1}} - 
         \sqrt{\Delta}\,\mathcal{A}_{\varphi,r}\,\underbrace{\vartheta^{3}\wedge 
         \vartheta^{1}}_{\boldsymbol{*}\vartheta^{2}} \right].
        \label{strau:60}
\end{eqnarray}
$\mathcal{A}_{\varphi}$ is explicitly (use (\ref{strau:59}), (\ref{strau:57}), 
(\ref{strau:54}), and (\ref{strau:51}))
\begin{displaymath}
        \mathcal{A}_{\varphi} =         
        \frac{1}{2}\,B_{0}\,\frac{\Sigma^{2}}{\rho^{2}}\,\sin^{2}\!\vartheta 
        \left(1-4a^{2}\,\frac{Mr}{\Sigma^{2}}\right).
\end{displaymath}
We write this as
\begin{equation}
        \mathcal{A}_{\varphi} = \frac{1}{2}\,B_{0}\,X, \qquad X = 
        \frac{\sin^{2}\!\vartheta}{\rho^{2}}\,(\Sigma^{2}-4a^{2}Mr).
        \label{strau:61}
\end{equation}
With this notation, (\ref{strau:60}) reads
\begin{equation}
        \mathcal{B}= \frac{B_{0}}{2\Sigma \sin\vartheta 
        }\,\left[X_{,\vartheta}\boldsymbol{*}\vartheta^{r} 
        -\sqrt{\Delta}\,X_{,r}\boldsymbol{*}\vartheta^{\vartheta} \right].
        \label{strau:62}
\end{equation}
The corresponding vector field $\vec{B}$ is thus
\begin{equation}
        \vec{B} =  \frac{B_{0}}{2\Sigma \sin\vartheta 
        }\,\left[X_{,\vartheta}\,\vec{e}_{r} 
        -\sqrt{\Delta}\,X_{,r}\,\vec{e}_{\vartheta} \right].
        \label{strau:63}
\end{equation}

\begin{figure}[htbp]
        \centerline{\includegraphics[width=10cm, height=10cm, angle=-1]{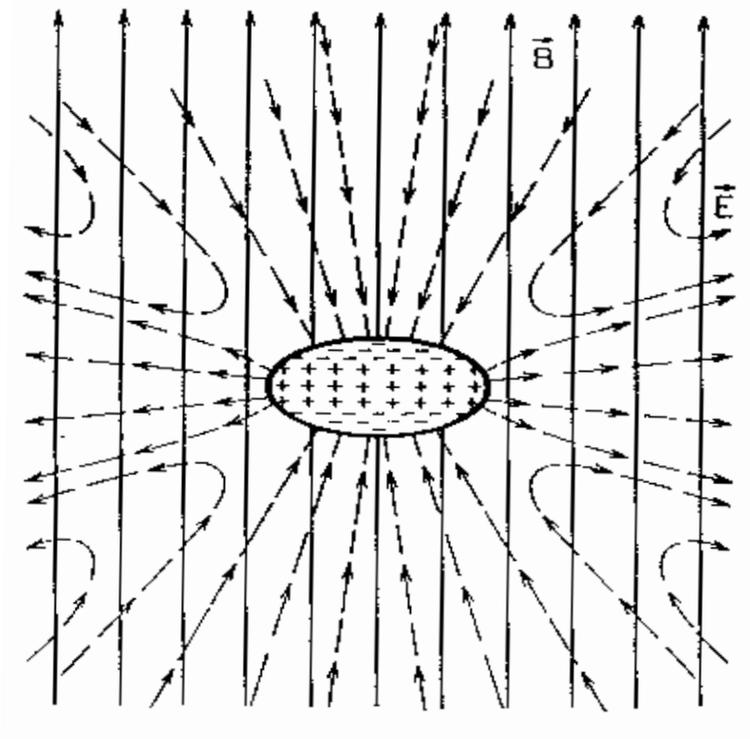}}
        \caption{Field lines of $\vec{E}$ (from Ref. [1]).}
        \protect\label{strau:Fig.3.1}
\end{figure}

For $\mathcal{E}$ we have, with $\beta = -\omega\,\partial_{\varphi}$,
\begin{equation}
        \alpha\,\mathcal{E}= -\mathbf{d}\phi + i_{\beta}\,\mathbf{d}\mathcal{A} = -\mathbf{d}\phi + 
        \omega\,\mathbf{d}\mathcal{A}_{\varphi}.
        \label{strau:64}
\end{equation}
>From this one finds quickly
\begin{eqnarray}
        \vec{E} & = & -\frac{B_{0}\,a\Sigma}{\rho^{2}} 
        \left\{\left[\frac{\partial\alpha^{2}}{\partial r} + 
        \frac{M\sin^{2}\!\vartheta}{\rho^{2}}\,(\Sigma^{2}-4a^{2}Mr)
        \frac{\partial}{\partial 
        r}\left(\frac{r}{\Sigma^{2}}\right)\right]\vec{e}_{r}\right.
        \nonumber \\
         &  & + \left. \frac{1}{\sqrt{\Delta}}\left[\frac{\partial\alpha^{2}}{\partial 
         \vartheta} + r\,
        \frac{M\sin^{2}\!\vartheta}{\rho^{2}}\,(\Sigma^{2}-4a^{2}Mr)
        \frac{\partial}{\partial\vartheta}\left(\frac{1}{\Sigma^{2}}\right)\right]
        \vec{e}_{\vartheta} \right\}.
        \label{strau:65}
\end{eqnarray}
The field lines of $\vec{E}$ are shown in Fig. \ref{strau:Fig.3.1}.

It is of interest to work out the magnetic flux through the equator 
of the BH, i.e.,
\begin{equation}
        \Phi = \int_{\textrm{upper h.}}\mathcal{B} = 
        \int_{\textrm{equator}}\mathcal{A} = 2\pi\,\mathcal{A}_{\varphi} 
        \left|_{\textrm{equator}}\right..
        \label{strau:66}
\end{equation}
Specializing (\ref{strau:61}) to $\vartheta = \pi/2$ and $r=r_{H}$ gives 
(the horizon is located at $\Delta=0$)
\begin{equation}
        \Phi = 4\pi B_{0}\,M(r_{H}-M) =4\pi B_{0}\,M\sqrt{M^{2}-a^{2}}.
        \label{strau:67}
\end{equation}
Here, we have used
\begin{equation}
        r_{H} = M + \sqrt{M^{2}-a^{2}}.
        \label{strau:68}
\end{equation}
Note that this vanishes for an extremal BH ($a=M$). Generically one 
has, as expected, $\Phi \approx \pi r_{H}^{2}B_{0}.$

\section{The Horizon as a Conducting Membrane}
As long as one is not interested in fine details very close to the 
horizon, one can regard the boundary conditions implied by the horizon 
as arising from physical properties of a fictitious membrane. This 
membrane is thought of as endowed with surface charge density 
$\sigma_{H}$, surface current $\vec{\mathcal{J}}_{H}$, and surface 
resistivity $R_{H}$. This idea was first pursued by Damour \cite{4} 
and independently by Znajek \cite{5}. Later it was further developed 
by Thorne and Mac\,Donald \cite{6} and other authors (see \cite{1} and 
references therein).

Below we give a much simplified derivation that the following surface 
properties for the parallel and perpendicular components hold:
\begin{eqnarray}
        \textrm{Gauss's law}: & E_{\perp} & \longrightarrow 4\pi 
        \,\sigma_{H},\nonumber  \\
        \textrm{Amp\`{e}re's law}: & \alpha \,\vec{B}_{\parallel} & 
        \longrightarrow \vec{B}_{H} = 4\pi\,\vec{\mathcal{J}}_{H}\times 
        \vec{n},  \nonumber  \\
        \textrm{charge conservation}: & \alpha\,j_{\perp} & \longrightarrow 
        -\partial_{t}\,\sigma_{H} -\, ^{(2)}\vec{\nabla}\cdot 
        \vec{\mathcal{J}}_{H},\nonumber  \\
        \textrm{Ohm's law}: & \alpha \,\vec{E}_{\parallel} & \longrightarrow 
        \vec{E}_{H} = R_{H}\,\vec{\mathcal{J}}_{H}.
        \label{strau:69}
\end{eqnarray}
The surface resistivity $R_{H}$ turns out to be equal to the vacuum 
impedance
\begin{equation}
        R_{H} = 4\pi = 377\textrm{ Ohm}.
        \label{strau:70}
\end{equation}
For the horizon fields $\vec{E}_{H}$, $\vec{B}_{H}$ we have, 
therefore,
\begin{equation}
        \vec{B}_{H} = \vec{E}_{H} \times \vec{n},
        \label{strau:71}
\end{equation}
as for a plane wave in vacuum.

Toward the horizon the FIDOs move relative to freely falling 
observers, say, more and more rapidly, approaching the velocity of 
light. Mathematically, the tetrad $\{\theta^{\mu}\}$ becomes singular 
at the horizon, and therefore the components of $F$ relative to 
$\{\theta^{\mu}\}$ are ill-behaved. The 2-form $F$ should, of course, 
remain regular and the laws (\ref{strau:69}) and (\ref{strau:70}) are just an 
expression of this requirement. For illustration, we demonstrate this 
first for a Schwarzschild BH, and generalize afterwards the argument 
in a simple manner.

\begin{center}
        \large{\textbf{(a) Derivation for a Schwarzschild BH}}
\end{center}

The procedure is simple: we pass to a regular, not necessarily 
orthonormal tetrad. The angular part $\theta^{2}=r\,d\vartheta$, 
$\theta^{3}=r\sin\!\vartheta\, d\varphi$ is kept, but instead of 
$\theta^{0}=\alpha\,dt$, $\theta^{1}=\alpha^{-1}\,dr$ we use 
$d\bar{t}$, where $\bar{t}$ is the Eddington-Finkelstein time 
coordinate
\begin{equation}
        \bar{t} = t + 2M\,\ln \left(\frac{r}{2M}-1\right) \longrightarrow 
        d\bar{t} = dt + \alpha^{-2}\,\frac{2M}{r}\,dr \qquad (r>2M).
        \label{strau:72}
\end{equation}
Since the Eddington-Finkelstein coordinates are regular at the 
horizon, the same is true for the basis 
$\{d\bar{t},\,dr,\,\theta^{2},\,\theta^{3}\}$. We have the relations
\begin{eqnarray}
        \theta^{0} & = & \alpha\,d\bar{t} - \frac{1}{\alpha}\,\frac{2M}{r}\,dr,\nonumber  \\
        \theta^{1} & = & \frac{1}{\alpha}\,dr,
        \label{strau:73}
\end{eqnarray}
which allow us to rewrite the decomposition (\ref{strau:6}) as follows
\begin{eqnarray*}
        F & = & E_{1}\,\theta^{1}\wedge \theta^{0} + \ldots + 
        B_{3}\,\theta^{1}\wedge \theta^{2} + \ldots  \\
         & = & E_{1}\,dr\wedge d\bar{t} + E_{2}\,\left(\alpha\,\theta^{2}\wedge 
         d\bar{t} - \frac{1}{\alpha}\,\frac{2M}{r}\,\theta^{2}\wedge dr\right) + 
         E_{3}\,(\ldots)  \\
        & & +B_{3}\,\frac{1}{\alpha}\,dr\wedge \theta^{2} -B_{2}\,\frac{1}{\alpha}\,dr\wedge \theta^{3}
        + B_{1}\,\theta^{2} \wedge \theta^{3}
\end{eqnarray*}
or
\begin{eqnarray}
        F & = & E_{1}\,dr \wedge d\bar{t} + \alpha \,E_{2}\,\theta^{2} \wedge 
        d\bar{t}  + \alpha \,E_{3}\,\theta^{3} \wedge d\bar{t} + 
        B_{1}\,\theta^{2} \wedge \theta^{3} \nonumber \\
         &  & + \frac{1}{\alpha}\left(B_{3} + \frac{2M}{r}\,E_{2}\right)\,dr 
         \wedge \theta^{2} + \frac{1}{\alpha}\left(-B_{2} + 
         \frac{2M}{r}\,E_{3}\right)\,dr 
         \wedge \theta^{3}.
        \label{strau:74}
\end{eqnarray}
The regularity of the coefficients in this expansion implies the 
following behavior when the horizon is approached $(\alpha \downarrow 
0)$:
\begin{eqnarray}
        (i) &\textrm{radial components:}  & E_{1},\:B_{1}=\mathcal{O}(1),
        \nonumber  \\
        (ii) & \textrm{tangent components:}&            
        E_{2},\:E_{3},\:B_{2},\:B_{3}=\mathcal{O}\left(\frac{1}{\alpha}\right),
        \nonumber \\
        (iii) & E_{2}+B_{3},\;E_{3}-B_{2} & = \mathcal{O}(\alpha).
        \label{strau:75}
\end{eqnarray}
This shows that
\begin{eqnarray}
    \alpha\,\vec{E}_{\parallel}& = & \vec{n}\times 
        \alpha\,\vec{B}_{\parallel}+\mathcal{O}(\alpha^{2}), 
        \qquad (\vec{n}:=\vec{e}_{r}), \nonumber \\
        \alpha\,\vec{B}_{\parallel} & = & -\vec{n}\times \alpha\,
        \vec{E}_{\parallel}+\mathcal{O} (\alpha^{2}),  \\
        E_{n}\equiv E_{\perp}, &  & B_{n}\equiv B_{\perp}\:\textrm{remain 
        finite}.
        \label{strau:76}
\end{eqnarray}
This is basically already what was claimed in (\ref{strau:69}) and 
(\ref{strau:70}). It is useful to introduce the \emph{stretched horizon} 
$\mathcal{H}^{s} = \{\alpha=\alpha_{H}\ll 1\}$ which is arbitrary 
close to the event horizon. Let
\begin{equation}
        \vec{E}_{H}:={(\alpha\,\vec{E}_{\parallel})}_{{\alpha}_{H}},\qquad 
        \vec{B}_{H}:= {(\alpha\,\vec{B}_{\parallel})}_{{\alpha}_{H}}
        \label{strau:77}
\end{equation}
and as above
\begin{equation}
        E_{n}:= \vec{E}\cdot \vec{n},\qquad B_{n}:= \vec{B}\cdot \vec{n}.
        \label{strau:78}
\end{equation}
These components remain \emph{finite} for $\alpha_{H} \downarrow 0$ 
and we have up to $\mathcal{O}(\alpha^{2}_{H})$
\begin{equation}
        \vec{E}_{H} = \vec{n}\times \vec{B}_{H}, \qquad 
        \vec{B}_{H}=-\vec{n}\times \vec{E}_{H}.
        \label{strau:79}
\end{equation}
The surface charge density $\sigma_{H}$ and the surface current 
density $\vec{\mathcal{J}}_{H}$ on $\mathcal{H}^{s}$ are defined by
\begin{equation}
        \sigma_{H}:=\left(\frac{E_{n}}{4\pi}\right) _{\mathcal{H}^{s}}, 
        \qquad  \vec{B}_{H}  =: \left( 4\pi\,\vec{\mathcal{J}}_{H}\times \vec{n} 
        \right) _{\mathcal{H}^{s}}.
        \label{strau:80}
\end{equation}
The second equation and (\ref{strau:50}) imply Ohm's law in (\ref{strau:69})
\begin{equation}
        \vec{\mathcal{J}}_{H} = \frac{1}{R_{H}}\,\vec{E}_{H}, \qquad 
        R_{H}=4\pi.
        \label{strau:81}
\end{equation}

Finally, we make use of the Amp\`{e}re-Maxwell law (\ref{strau:39}) (for 
$\vec{\beta}=0$):
\begin{displaymath}
        \partial_{t}\,\vec{E} = \vec{\nabla} \times (\alpha\,\vec{B}) - 4\pi 
        \,\alpha\,\vec{j}.
\end{displaymath}
The normal component on $\mathcal{H}^{s}$ is
\begin{displaymath}
        \partial_{t}\,E_{n} = \left. \left[\vec{\nabla} \times 
        (\alpha\,\vec{B})\right]_{n} - 4\pi\,\alpha\,\vec{j}_{n} \; \right |
        _{\mathcal{H}^{s}}.
\end{displaymath}
Using
\begin{displaymath}
        \left.\left[\vec{\nabla} \times (\alpha\,\vec{B})\right]_{n} \; 
        \right|_{\mathcal{H}^{s}} = \left.      \left[\vec{\nabla} \times (4\pi\,\vec{\mathcal{J}}_{H}\times 
        \vec{n})\right]_{n} \; \right|_{\mathcal{H}^{s}} = -\,4\pi\,^{(2)}
        \vec{\nabla}\cdot \vec{\mathcal{J}}_{H},
\end{displaymath}
where $^{(2)}\vec{\nabla}$ denotes the induced 
covariant derivation on $\mathcal{H}^{s}$, we obtain
\begin{equation}
        \partial_{t}\,\sigma_{H} +\, ^{(2)}\vec{\nabla}\cdot 
        \vec{\mathcal{J}}_{H} + (\alpha\,j_{n})_{\mathcal{H}^{s}}=0,
        \label{strau:82}
\end{equation}
as an expression of charge conservation.

This completes the derivation of (\ref{strau:69}) and (\ref{strau:70}) for the 
Schwarzschild BH. Next, we generalize the discussion to an arbitrary 
static BH.

\begin{center}
        \large{\textbf{(b) Derivation for static BH}}
\end{center}

We introduce first a parametrization of the exterior metric which was 
used also in Israel's famous proof of the uniqueness theorem for the 
Schwarzschild BH. 

The starting point is (\ref{strau:3}) for $\beta=0$, i.e.,
\begin{equation}
        ^{(4)}\boldsymbol{g}=-\alpha^{2}dt^{2}+\boldsymbol{g}, \quad \alpha 
        \textrm{ and } \boldsymbol{g} \textrm{ independent of }t.
        \label{strau:83}
\end{equation}
Note that $\alpha^{2}=-(k|k)\ge 0,\: k=\partial_{t}$, and that the 
horizon has to be at $\alpha=0$. We assume that the lapse function has 
no critical point, $d\alpha \neq 0$. The absolute space $\Sigma$ is 
then foliated by the leaves $\{\alpha=\textrm{const}\}$. The function
\begin{equation}
        \rho := (d\alpha|d\alpha)^{-\frac{1}{2}}
        \label{strau:84}
\end{equation}
is then positive on $\Sigma$.

Now we introduce adapted coordinates on $\Sigma$. Consider in any 
point $p \in \Sigma$ the 1-dimensional subspace of $T_{p}(\Sigma)$ 
perpendicular to the tangent space of the leave 
$\{\alpha=\textrm{const}\}$ through $p$. This defines a 1-dimensional 
distribution which is, of course, involutive (integrable). The 
Frobenius theorem then tells us that we can introduce coordinates 
$\{x^{i}\}$ on $\Sigma$, such that $x^{A}$ ($A=2,\,3)$ are constant 
along the integral curves of the distribution. For $x^{1}$ we can 
choose the lapse function, and thus obtain
\begin{equation}
        \boldsymbol{g}=\rho^{2}d\alpha^{2}+\tilde{\boldsymbol{g}}, \quad 
        \tilde{\boldsymbol{g}}=\tilde{g}_{AB}dx^{A}dx^{B}.
        \label{strau:85}
\end{equation}
Here, $\rho$ and $\tilde{g}_{AB}$ depend in general on all three 
coordinates $x^{1}=\alpha,\,x^{A}$.

We also need the surface gravity $\kappa$ on the horizon. A useful 
formula is (see, e.g., \cite{7})
\begin{equation}
        \kappa^{2} = -\frac{1}{4}(dk|dk)\big| _{H}.
        \label{strau:86}
\end{equation}
Now, $k=-\alpha^{2}dt$, $dk=-2\alpha\,d\alpha \wedge dt$, whence
\begin{equation}
        \kappa = \frac{1}{\rho_{H}}.
        \label{strau:87}
\end{equation}
>From the zeroth law of BH physics we know that $\kappa = 
\textrm{const}$. Since we want to assume a regular Killing horizon 
(generated by $k$), we conclude
\begin{equation}
                0 < \rho_{H} < \infty ,\qquad \rho_{H} = \textrm{const}.
        \label{strau:88}
\end{equation}
Combining (\ref{strau:83}) with (\ref{strau:85}) we have outside the horizon
\begin{equation}
        ^{(4)}\boldsymbol{g} = -N\,dt^{2} + \frac{\rho^{2}}{4N}\,dN^{2} + 
        \tilde{\boldsymbol{g}}, \qquad N:=\alpha^{2}. 
        \label{strau:89}      
\end{equation}
The natural FIDO tetrad is
\begin{eqnarray}
        \theta^{0}& = &\sqrt{N}\,dt ,\nonumber  \\
        \theta^{1}& = &\frac{\rho}{2\sqrt{N}}\,dN ,     \nonumber  \\
        \theta^{A}& &  (A=2,3): \quad \textrm{orthonormal 2-bein for }\tilde{\boldsymbol{g}}.
        \label{strau:90}
\end{eqnarray}
This becomes again singular at the horizon ($N=0$).

Now we imitate what we did for the Schwarzschild BH. We search for a basis 
of 1-forms which is well-defined in the neighborhood of the horizon. 
Guided by (\ref{strau:72}) we introduce
\begin{equation}
        \bar{\theta}^{t} \equiv dt + \frac{\rho}{2N}\,dN
        \label{strau:91}
\end{equation}
and rewrite (\ref{strau:89})
\begin{displaymath}
        ^{(4)}\boldsymbol{g} = 
        -N\,(\bar{\theta}^{t})^{2}+\rho\,\bar{\theta}^{t}\,dN+\tilde{\boldsymbol{g}}.
\end{displaymath}
Thanks to (\ref{strau:88}) we conclude that
\begin{equation}
        \left\{\bar{\theta}^{t},\, dN,\, \theta^{A}\:(A=2,3)\right\}
        \label{strau:92}
\end{equation}
remains a regular basis on the horizon. Outside the horizon we can 
express (\ref{strau:90}) in terms of this basis:
\begin{eqnarray}
        \theta^{0} &=& 
        \sqrt{N}\,\bar{\theta}^{t}-\frac{\rho}{2\sqrt{N}}\,dN ,
        \nonumber  \\
        \theta^{1} &=& \frac{\rho}{2\sqrt{N}}\,dN.
        \label{strau:93}
\end{eqnarray}

We can now proceed as in the derivation of (\ref{strau:74}), obtaining now
\begin{eqnarray}
                F & = & \frac{\rho\,E_{1}}{2}\,dN\wedge 
                \bar{\theta}^{t}+\sqrt{N}E_{2}\,\theta^{2}\!\wedge 
                \bar{\theta}^{t}+\sqrt{N}E_{3}\,\theta^{3}\!\wedge 
                \bar{\theta}^{t}+B_{1}\theta^{2}\!\wedge\theta^{3} \nonumber \\
                 &  & +\frac{\rho}{2\sqrt{N}}\,(E_{2}+B_{3})\,dN\wedge\theta^{2} + 
                 \frac{\rho}{2\sqrt{N}}\,(B_{2}-E_{3})\,dN\wedge \theta^{3}.
        \label{strau:94}
\end{eqnarray}
This implies again the limiting behavior (\ref{strau:75}) for the normal 
and parallel components of the electric and magnetic fields.

\begin{center}
        \large{\textbf{(c) Derivation for rotating BHs}}
\end{center}

Finally, I give a similar derivation for rotating BHs. (This was worked 
out in collaboration with Gerold Betschart in the course of his diploma 
work.)

I first need the Papapetrou parametrization of a stationary, 
axisymmetric BH. Since this will be treated in detail in the lectures 
by Markus Heusler \cite{7}, I can be brief.

Since the isometry group is $\boldsymbol{R} \times SO(2)$ we have two 
commuting Killing fields $k$ and $m$, say, which are tangent to the 
orbits belonging to the group action. We assume that $k$ and $m$ 
satisfy the Frobenius integrability conditions
\begin{equation}
        k \wedge m \wedge dk =0, \qquad k \wedge m \wedge dm =0.
        \label{strau:95}
\end{equation}
The Frobenius theorem then guarantees that the distribution of 
subspaces orthogonal to $k$ and $m$ is (locally) integrable. I recall 
that (\ref{strau:95}) is implied by the field equations for vacuum 
spacetimes and also for certain matter models (electromagnetic 
fields, ideal fluids, but not for Yang-Mills fields).

In this situation, spacetime is (locally) a product manifold, 
$M=\Sigma \times \Gamma$, where $\Sigma = \boldsymbol{R} \times 
SO(2)$ and $\Gamma$ is perpendicular to $\Sigma$. Thus the metric splits
\begin{equation}
        ^{(4)}\boldsymbol{g}=\boldsymbol{\sigma} + \boldsymbol{g}
        \label{strau:96}
\end{equation}
such that $\boldsymbol{\sigma}$ is an invariant 2-dimensional Lorentz 
metric on $\Sigma$, depending, however, on $y \in \Gamma$, and 
the fact that $(\Gamma,\boldsymbol{g})$ is a 2-dimensional Riemannian space.
In adapted coordinates
\begin{equation}
        x^{\mu}:\quad x^{0}=t, \quad x^{1}=\varphi \textrm{ for } \Sigma;\quad x^{2},\, x^{3} 
        \textrm{ for } \Gamma,
        \label{strau:97}
\end{equation}
we have
\begin{equation}
        k = \partial_{t}, \qquad m = \partial_{\varphi},
        \label{strau:98}
\end{equation}
and
\begin{equation}
        \boldsymbol{\sigma} = \sigma_{ab}\,dx^{a}dx^{b}, \qquad 
        \boldsymbol{g} = g_{ij}\,dx^{i}dx^{j},
        \label{strau:99}
\end{equation}
where $a,b=0,1$ and $i,j=2,3$. The metric functions $\sigma_{ab}$ and 
$g_{ij}$ depend only on the coordinates $x^{i}$ of $\Gamma$.

The following functions on $\Gamma$ have an invariant meaning
\begin{equation}
        -V:= (k|k), \qquad W:= (k|m), \qquad X:= (m|m),
        \label{strau:100}
\end{equation}
and we have
\begin{equation}
        \boldsymbol{\sigma} = -V\,dt^{2} + 2W\,dtd\varphi + X\,d\varphi^{2}.
        \label{strau:101}
\end{equation}
We use also
\begin{equation}
        \rho := \sqrt{VX+W^{2}} = \sqrt{-\sigma},\qquad A:= \frac{W}{X},
        \label{strau:102}
\end{equation}
in terms of which (\ref{strau:101}) takes the form
\begin{equation}
        \boldsymbol{\sigma} = -\frac{\rho^{2}}{X}\, dt^{2} + X(d\varphi + 
        A\,dt)^{2}.
        \label{strau:103}
\end{equation}

It turns out that the partial trace $R^{a}_{\:a}$ of the 
4-dimensional Ricci tensor is proportional to $^{(g)}\Delta \rho$ 
and $\rho$ is thus a harmonic function on $\Gamma$, whenever 
$R^{a}_{\:a}$ vanishes. This is of course the case for vacuum 
manifolds, but also for the Kerr-Newman solution, and in some other 
cases \cite{7}. With the help of the Riemann mapping theorem one can 
then show that $\rho$ is a well-defined coordinate on 
$(\Gamma,\boldsymbol{g})$ ($\rho$ has no critical points). It is then 
possible to introduce a second coordinate $z$, such that
\begin{equation}
        \boldsymbol{g} = \frac{1}{X}e^{2h}(d\rho^{2} + dz^{2}).
        \label{strau:104}
\end{equation}
In terms of $t$, $\varphi$, and the \emph{Weyl coordinates} $\rho$, 
$z$, $^{(4)}\boldsymbol{g}$ assumes the \emph{Papapetrou 
parametrization}
\begin{equation}        
        ^{(4)}\boldsymbol{g}=-\frac{\rho^{2}}{X}\,dt^{2}+X\,(d\varphi+A\,dt)^{2} + 
        \frac{e^{2h}}{X}(d\rho^{2}+dz^{2}).
        \label{strau:105}
\end{equation}
We emphasize once more, that the functions $X$, $A$, and $h$ depend 
only on the Weyl coordinates $\rho$ and $z$.

The weak rigidity theorem tells us that on the surface on which
\begin{equation}
        \xi=k+\Omega\,m ,\qquad \Omega =-A=-W/X,
        \label{strau:106}
\end{equation}
becomes null, $\Omega$ is constant and $\xi$ is a Killing field on 
this surface, denoted by $H[\xi]$ in what follows. In addition, $H[\xi]$ 
is a stationary null surface, in particular a Cauchy horizon. (For 
proofs, see \cite{7}.)

Note that $\xi$, as a 1-form, can be expressed as
\begin{equation}
        \xi = -\frac{\rho^{2}}{X}\,dt.
        \label{strau:107}
\end{equation}

One also knows that $X$ is well-behaved on the horizon
\begin{equation}
        X = \mathcal{O}(1), \qquad X^{-1} = \mathcal{O}(1) \quad \textrm{on}\; H[\xi]
        \label{strau:108}
\end{equation}
(see \cite{8}). Since $(\xi|\xi)=-\rho^{2}/X$ the horizon is at 
$\rho = 0$ (as is well-known from the Kerr solution).

Although $\xi$ is not a Killing field, the weak rigidity theorem 
implies that the surface gravity is still given by
\begin{equation}
        \kappa^{2} = -\frac{1}{4}(d\xi|d\xi)\big|_{H[\xi]}.
        \label{strau:109}
\end{equation}
(A priori, the formula holds for $l:=k+\Omega_{H}\,m$, 
$\Omega_{H}=\Omega|_{H[\xi]}$, but $d\Omega =0$ on $H[\xi]$.) From 
(\ref{strau:107}) we obtain
\begin{displaymath}
                d\xi = -\frac{2\rho}{X}\, d\rho \wedge dt - \rho^{2}\, 
        d\left(\frac{1}{X}\right)\wedge dt.
\end{displaymath}
Using the scalar products $(d\rho|d\rho)=Xe^{-2h}$, $(dt|dt) = -X/\rho^{2}$, 
$(d\rho|dt)=0$ gives thus, together with the zeroth law,
\begin{equation}
        \kappa = {e^{-h}}\big|_{H[\xi]} = \textrm{const} \neq 0.
        \label{strau:110}
\end{equation}
The reader should verify  that this gives the correct result for the 
Kerr solution
\begin{equation}
        \kappa = \frac{r_{H}-M}{2Mr_{H}}.
        \label{strau:111}
\end{equation}

After these preparations, our argument proceeds similarly as in (b). 
The natural FIDO tetrad $\{\theta^{\mu}\}$ for (\ref{strau:105}) is 
$(N:=\rho^{2})$
\begin{equation}
        \theta^{0} = \sqrt{\frac{N}{X}}\,dt,\quad \theta^{1} = \frac{e^{h}}{2\sqrt{XN}}\,dN,
        \quad \theta^{2} = \frac{e^{h}}{\sqrt{X}}\,dz ,\quad \theta^{3} =\sqrt{X}(d\varphi + A\,dt). 
        \label{strau:112}
\end{equation}
Clearly, $\theta^{0}$ and $\theta^{1}$ are again ill-defined on the 
horizon $N=0$. We therefore pass over to a new basis
\begin{equation}
        \bar{\theta}^{t}:=dt + \frac{e^{h}}{2N}\,dN, \qquad 
        \bar{\theta}^{\varphi}:= d\varphi + \Omega\,\frac{e^{h}}{2N}\,dN,
        \label{strau:113}
\end{equation}
together with $dN$ and $dz$. In order to check whether this new basis 
remains valid when the horizon is approached, we express the metric 
(\ref{strau:105}) in terms of it. Since
\begin{equation}
        \theta^{0} = \sqrt{\frac{N}{X}}\,\bar{\theta}^{t} - 
        \frac{e^{h}}{2\sqrt{XN}}\,dN , \qquad \theta^{3} = 
        \sqrt{X}(\bar{\theta}^{\varphi} + A\,\bar{\theta}^{t}), 
        \label{strau:114}
\end{equation}
we find readily
\begin{equation}
        ^{(4)}\boldsymbol{g}= -V(\bar{\theta}^{t})^{2} + 
         2W\,\bar{\theta}^{t}\bar{\theta}^{\varphi} + 
         X(\bar{\theta}^{\varphi})^{2} + \frac{e^{h}}{X}\,\bar{\theta}^{t}dN + 
         \frac{e^{2h}}{X}\,dz^{2}.    
        \label{strau:115}
\end{equation}
The determinant of the metric coefficients in this expression is $ 
\frac{e^{4h}}{4X^{2}}$ and thus remains regular, thanks to (\ref{strau:107}) 
and (\ref{strau:109}). Since we postulate a regular horizon, it is then 
clear that $\{\bar{\theta}^{t},\,\bar{\theta}^{\varphi},\,dN,\,dz\}$ 
indeed forms a well-defined basis also on the horizon.

It is now straightforward to express the Maxwell 2-form $F$ in terms 
of this basis. Instead of (\ref{strau:95}) we obtain now
\begin{eqnarray}
        F & = & \left\{ E_{1}\,\frac{e^{h}}{2X} + 
         \frac{Ae^{h}}{2\sqrt{N}}(E_{3}-B_{2})\right\}dN\wedge 
         \bar{\theta}^{t} + E_{3}\sqrt{N}\,\bar{\theta}^{\varphi}\wedge \bar{\theta}^{t} \nonumber \\
         & & \left\{E_{2}e^{h}\sqrt{\frac{N}{X}} + B_{1}e^{h}A\right\}dz\wedge 
         \bar{\theta}^{t} + \frac{e^{h}}{2\sqrt{N}}(B_{2}-E_{3})\,dN\!\wedge \bar{\theta}^{\varphi}
         \nonumber \\
         & & +B_{1}e^{h}\,dz\wedge\bar{\theta}^{\varphi} - 
         \frac{e^{2h}}{2X\sqrt{N}}(E_{2}+B_{3})\,dz\wedge dN .
        \label{strau:116}
\end{eqnarray}
Since $\alpha:=\sqrt{\frac{N}{X}}$ is the lapse function, we obtain 
once more the limiting behavior (\ref{strau:75}), and thus the basic membrane 
laws (\ref{strau:69}) and (\ref{strau:70}).

\section{Magnetic Energy Extraction from a Black Hole}

As an interesting, and possibly astrophysically important application 
of our basic laws (\ref{strau:38}) and (\ref{strau:69}) I show now that it is 
possible, in principle, to extract the rotational energy of a BH with 
the help of external magnetic fields. In the next section, we will 
work out some of the details for an ideal gedanken experiment. This 
will serve as a preparation for an understanding of the 
Blandford-Znajek process.

Our starting point is Faraday's induction law (\ref{strau:25}) in integral 
form, which we write down once more
\begin{equation}
        \textrm{EMF}(\mathcal{C}) = 
        -\frac{d}{dt}\int_{\mathcal{A}}^{}\mathcal{B} + 
        \oint_{\mathcal{C}}^{}i_{\beta}\,\mathcal{B}. 
        \label{strau:5.1}
\end{equation}
For stationary situations this reduces to
\begin{equation}
        \textrm{EMF}(\mathcal{C}) =  
        \oint_{\mathcal{C}}^{}i_{\beta}\,\mathcal{B}. 
        \label{strau:5.2}
\end{equation}

\begin{figure}[htbp]
        \centerline{\includegraphics[width= 8cm, height=6cm ]{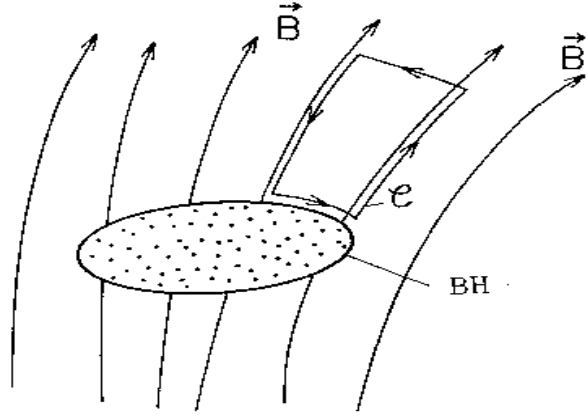}}
        \caption{Arrangement for eq. (\ref{strau:5.2}).}
        \protect\label{strau:Fig.5.1}
\end{figure}

In Fig. \ref{strau:Fig.5.1} we consider a stationary rotating BH in an external 
magnetic field (like in \S 3). The integral in (\ref{strau:5.2}) along the 
field lines gives no contribution and far away $\beta$ drops rapidly 
($\sim r^{-2}$). Thus, there remains only the contribution from the 
horizon ($\mathcal{C}_{H}$) of the path $\mathcal{C}$ in Fig. 
\ref{strau:Fig.5.1}:
\begin{equation}
        \textrm{EMF} = \int_{\mathcal{C}_{H}}^{}i_{\beta_{H}}\,\mathcal{B}, \quad        
        \beta_{H}=-\Omega_{H}\,\partial_{\varphi}= 
        -\Omega_{H}\,\tilde{\omega}_{H}\,e_{\varphi}
        \label{strau:5.3}
\end{equation}
(only the normal component $\vec{B}_{\perp}$ contributes). I recall 
that $\Omega_{H}=a(2Mr_{H})^{-1},\ r_{H}=M+\sqrt{M^{2}-a^{2}}$.

We shall show in section 6 that it is possible to construct a 
generator such that (with optimal impedance matching) a \emph{maximal 
extraction rate} equal to
\begin{equation}
        \frac{1}{4}\,\frac{(\textrm{EMF})^{2}}{R(\mathcal{C}_{H})}
        \label{strau:5.4}
\end{equation}
becomes possible, where $R(\mathcal{C}_{H})$ is the horizon 
(``internal'') resistance
\begin{equation}
        R(\mathcal{C}_{H}) = 
        \int_{\mathcal{C}_{H}}R_{H}\,\frac{dl}{2\pi\,\tilde{\omega}}, \qquad 
        R_{H} = 377\textrm{ Ohm.}
        \label{strau:5.5}
\end{equation}
(It should be known from electrodynamics, that we have to divide the 
surface resistivity $R_{H}$ by the length of the cross section through 
which the current is flowing.) We shall also show that an equal amount 
of energy is dissipated by ohmic heating at the (stretched) horizon.

\begin{figure}[htbp]
        \centerline{\includegraphics[width=8cm , height=6cm ]{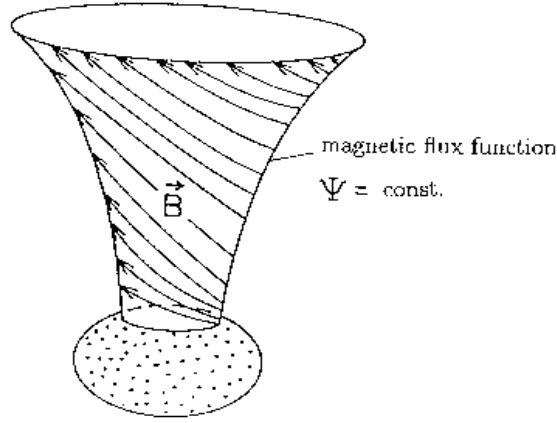}}
        \caption{Axisymmetric magnetic field. The total flux inside the 
        magnetic surface defines the flux function $\Psi$.}
        \protect\label{strau:Fig.5.2}
\end{figure}

Let us work this out for the special case of an axisymmetric field: 
$L_{\partial_{\varphi}}\,\mathcal{B}=0\leftrightarrow 
\mathbf{d}i_{\partial_{\varphi}}\,\mathcal{B}=0$, whence
\begin{equation}
        i_{\partial_{\varphi}}\,\mathcal{B}= -\frac{\mathbf{d}\Psi}{2\pi}.
        \label{strau:5.6}
\end{equation}
>From this we conclude that $\mathcal{B}$ can be expressed in terms of 
two potentials $\Psi$ and $g$,
\begin{equation}
        \mathcal{B} =\underbrace{ \frac{1}{2\pi}\, \mathbf{d}\Psi \wedge \mathbf{d}\varphi}_{\textrm
        {poloidal part}} + \underbrace{g\boldsymbol{*}\mathbf{d}\varphi}_{\textrm{toroidal part}},
        \label{strau:5.7}
\end{equation}
both of which can be taken to be independent of $\varphi$. This is 
equivalent to the vector formulae
\begin{equation}
        \vec{B}^{\textrm{pol}} =\frac{1}{2\pi\,\tilde{\omega}}\, 
        \vec{\nabla}\Psi \times \vec{e}_{\varphi}, \qquad 
        \vec{B}^{\textrm{tor}} = \frac{g}{\tilde{\omega}}\,\vec{e}_{\varphi}
        \label{strau:5.8}
\end{equation}
(Exercise). $\Psi$ is the magnetic flux function (see Fig. \ref{strau:Fig.5.2}), 
because the poloidal flux inside a tube $\{\Psi = \textrm{const}\}$ is
\begin{equation}
        \int \mathcal{B} = \int \mathbf{d}\Psi = \Psi, \qquad \Psi(0)=0.
        \label{strau:5.9}
\end{equation}
$\Psi$ is constant along magnetic field lines, as should be clear 
from Fig. \ref{strau:Fig.5.2}. Formally, this comes about as follows:
\begin{eqnarray*}
        i_{\vec{B}}\,\mathbf{d}\Psi & = & \boldsymbol{*}(\boldsymbol{*}
        \mathbf{d}\Psi\wedge B) = \boldsymbol{*}(\mathbf{d}\Psi \wedge \,\boldsymbol{*}B) = 
        \boldsymbol{*}(\mathbf{d}\Psi \wedge \mathcal{B}) \\
         & \stackrel{(\ref{strau:5.7})}{=} & g\boldsymbol{*}(\mathbf{d}\Psi \wedge \,
         \boldsymbol{*}\mathbf{d}\varphi)=0,
\end{eqnarray*}
thus $<\mathbf{d}\Psi,\vec{B}>=0$.

\begin{figure}[htbp]
        \centerline{\includegraphics[width=8cm , height=6cm ]{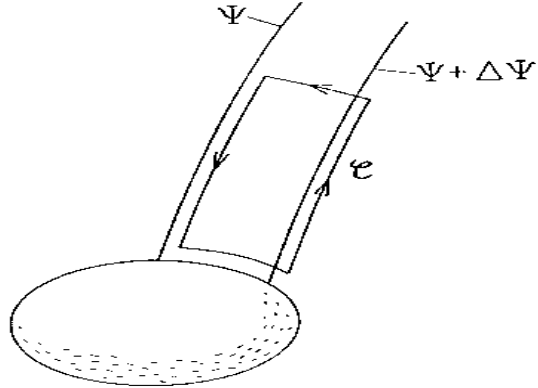}}
        \caption{Integration path for the voltage in eq. (\ref{strau:5.10}).}
        \protect\label{strau:Fig.5.3}
\end{figure}

For the closed path $\mathcal{C}$ in Fig. \ref{strau:Fig.5.3} the EMF is by 
(\ref{strau:5.3})
\begin{displaymath}
        \textrm{EMF} \equiv \triangle V = \int_{\mathcal{C}_{H}}^{}i_{\beta_{H}}\,\mathcal{B}
        \stackrel{(\ref{strau:5.6})}{=} -\left(-\frac{\Omega_{H}}{2\pi} \right)\,
        \int_{\mathcal{C}_{H}}\mathbf{d}\Psi,
\end{displaymath}
i.e.
\begin{equation}
        \triangle V = \frac{\Omega_{H}}{2\pi}\,\triangle \Psi.
        \label{strau:5.10}
\end{equation}

\begin{figure}[htbp]
        \centerline{\includegraphics[width=13cm , height=7cm ]{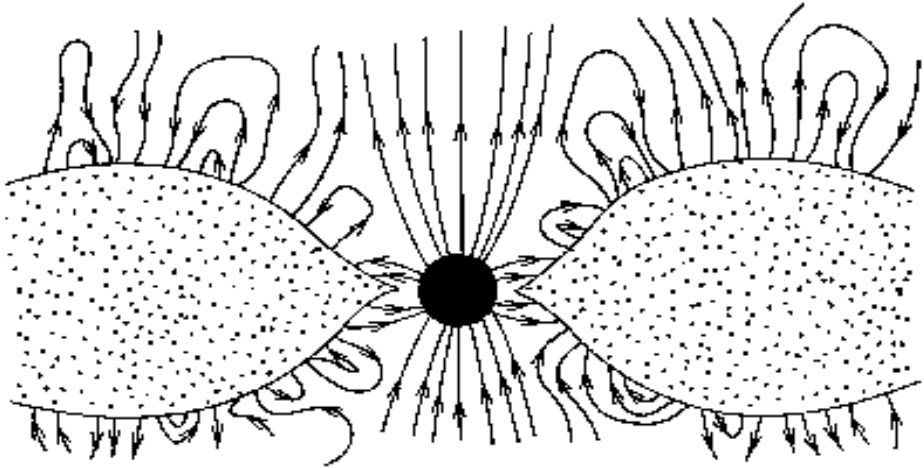}}
        \caption{Plausible structure of the magnetic field close to a 
        supermassive BH which is surrounded by an accretion disk (central 
        engine of a quasar) (adapted from Ref.\,[1]).}
        \protect\label{strau:Fig.5.4}
\end{figure}

The internal resistance (\ref{strau:5.5}) becomes
\begin{equation}
        \triangle R_{H} = R_{H}\, \frac{\triangle l}{2\pi\,\tilde{\omega}}.
        \label{strau:5.11}
\end{equation}
Since, on the other hand,
\begin{equation}
        \triangle \Psi = 2\pi\,\tilde{\omega}\,B_{\perp}\,\triangle l,
        \label{strau:5.12}
\end{equation}
we find, by eliminating $\triangle l$,
\begin{equation}
        \triangle R_{H} = R_{H}\, \frac{\triangle 
        \Psi}{4\pi^{2}\,\tilde{\omega}^{2}\,B_{\perp}}.
        \label{strau:5.13}
\end{equation}
Inserting (\ref{strau:5.10}) and (\ref{strau:5.13}) into the expression (\ref{strau:5.4}) 
for the maximal power output gives
\begin{equation}
        \frac{1}{4}\,\frac{(\triangle V)^{2}}{\triangle R_{H}} =        
        \frac{\Omega_{H}^{2}}{16\,\pi}\,\tilde{\omega}^{2}\,B_{\perp}\,
        \triangle \Psi .
        \label{strau:5.14}
\end{equation}
This, as well as (\ref{strau:5.10}), have to be integrated from the pole to 
some point north of the equator (see Fig. \ref{strau:Fig.5.4}). For the exact solution 
in \S 3 we know the result for the EMF, if we integrate up to the 
equator: From (\ref{strau:5.10}) and (\ref{strau:67}) we get
\begin{displaymath}
        \textrm{EMF} = \frac{1}{2\,\pi}\,\Omega_{H}\,4\pi\,B_{0}\,M\,(r_{H}-M)
\end{displaymath}
or $(\Omega_{H}=a/2Mr_{H})$
\begin{equation}
        \textrm{EMF}  = a\,B_{0}\,\frac{r_{H}-M}{r_{H}}\qquad 
        \left(r_{H}=M+\sqrt{M^{2}-a^{2}}\right).
        \label{strau:5.15}
\end{equation}

For a general situation, like in Fig. 6, we have roughly
\begin{displaymath}
        \Sigma\,\triangle \Psi=\Psi \sim B_{\perp}\,\pi\,r_{H}^{2}, \qquad
        \tilde{\omega}^{2} \sim \ <\tilde{\omega}^{2}>\  \sim \frac{r_{H}^{2}}{2},
\end{displaymath}
and we obtain, by (\ref{strau:5.14}), for the power output
\begin{eqnarray}
        P & \sim & 
        \frac{1}{128}\,\left(\frac{a}{M}\right)^{2}B_{\perp}^{2}\,r_{H}^{2}
        \nonumber\\
         & \sim & (10^{45}\,\textrm{erg/s})\,\left(\frac{a}{M}\right)^{2}        
         \left(\frac{M}{10^{9}\,M_{\odot}}\right)^{2}
         \left(\frac{B_{\perp}}{10^{4}\,G}\right)^{2}.
        \label{strau:5.16}
\end{eqnarray}
The total EMF is $(V=\Sigma\,\triangle V)$
\begin{equation}
        V \sim \frac{1}{2\pi}\,\Omega_{H}\,\Psi \sim \frac{1}{2\pi}\,
        \frac{a}{2Mr_{H}}\,B_{\perp}\,\pi\,r_{H}^{2} \simeq \frac{1}{2}
        \left(\frac{a}{M}\right)MB_{\perp}
        \label{strau:5.17}
\end{equation}
(compare this with (\ref{strau:5.15})). Numerically we find
\begin{equation}
        V \sim (10^{20}\,\textrm{Volt}) \left(\frac{a}{M}\right)\,
        \frac{M}{10^{9}\,M_{\odot}}\,\frac{B_{\perp}}{10^{4}\,G}.
        \label{strau:5.18}
\end{equation}
For reasonable astrophysical parameters we obtain magnetospheric 
voltages $V$ $ \sim 10^{20}$ Volts and power output of the magnitude $\sim
10^{45}\textrm{ erg/s}$. This power is what one observes typically in active 
galactic nuclei, and the voltage is comparable to the highest cosmic 
ray energies that have been detected.

Note, however, that for a realistic astrophysical situation there is 
plasma outside the BH and it is, therefore, not clear, how the horizon 
voltage (\ref{strau:5.18}) is used in accelerating particles to very high 
energies.

Let us estimate at this point the characteristic magnetic field 
strength than can be expected outside a supermassive BH. A measure 
for this is the field strength $B_{E}$ for which the energy density 
$B_{E}^{2}/8\pi$ is equal to the radiation energy density $u_{E}$ 
corresponding to the Eddington luminosity
\begin{equation}
        L_{E} = \frac{4\pi M_{H}m_{p}c}{\sigma_{T}} = 1.3 \times 
        10^{38}\,(\textrm{erg/s})\,\frac{M}{M_{\odot}}.
        \label{strau:5.19}
\end{equation}
The relation between $L_{E}$ and $u_{E}$ is
\begin{equation}
        L_{E} = 4\pi\,r_{g}^{2}\,\frac{c}{4}\,u_{E} = \pi\,r_{g}^{2}\,c\,u_{E}
        \qquad \left(r_{g}=\frac{G\,M}{c^{2}}\right).
        \label{strau:5.20}
\end{equation}
Thus
\begin{equation}
        \frac{1}{8\,\pi}\,B_{E}^{2} = 
        \frac{4\,m_{p}c^{2}}{\sigma_{T}\,r_{g}},
        \label{strau:5.21}
\end{equation}
giving $(M_{H,8} \equiv M_{H}/10^{8}\,M_{\odot})$
\begin{equation}
        B_{E} = 1.2 \times 10^{5}\,M_{H,8}^{-1/2}\ \textrm{Gauss}.
        \label{strau:5.22}
\end{equation}
For a BH with mass $\sim 10^{9}\ M_{\odot}$ inside an accretion disk 
acting as a dynamo, a characteristic field of about 1 Tesla ($10^{4}$ Gauss)
is thus quite reasonable.

\section{Rotating BH as a Current Generator}

\begin{figure}[htbp]
        \centerline{\includegraphics[width=15cm , height=10cm ]{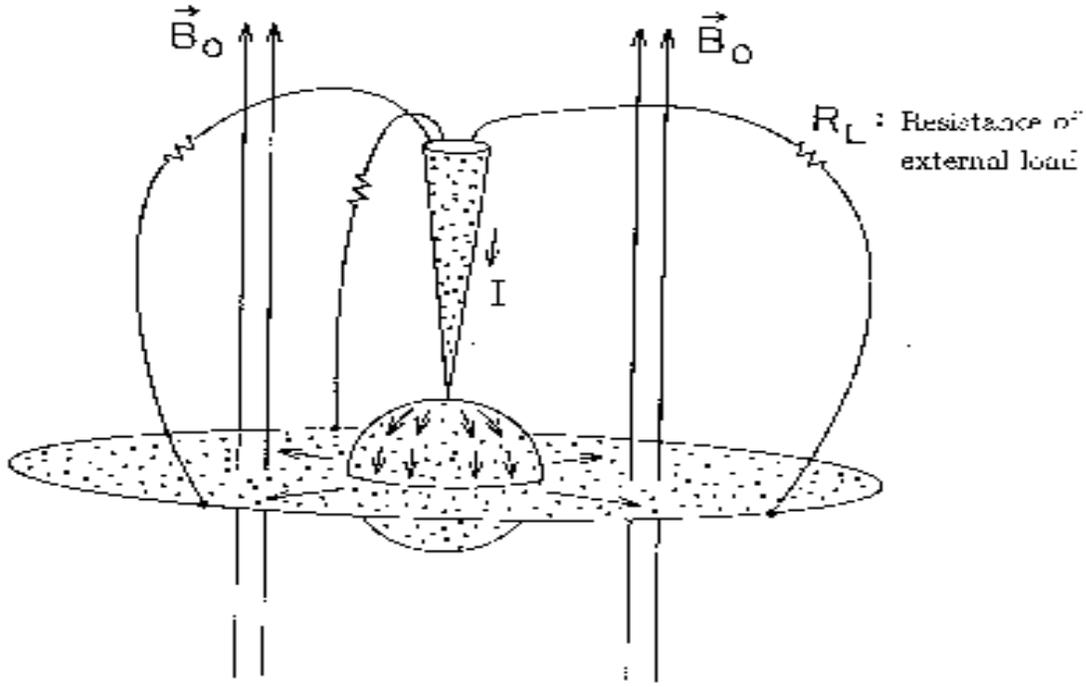}}
        \caption{An idealized current generator (adapted from [1]).}
        \protect\label{strau:Fig.6.1}
\end{figure}

Before we come to realistic possibilities of energy extraction, we 
analyze in detail an idealized arrangement, sketched in Fig. 
\ref{strau:Fig.6.1}.

We compute first the EMF around a closed path consisting of the 
following parts: We start from the equator of the (stretched) horizon 
along a perfectly conducting disk, which is supposed to rotate 
differentially in such a manner, that all its pieces are at rest 
relative to the FIDOs (they have thus zero angular momentum). From the 
boundary of the disk, the path continuous along a wire and through a 
resistive load ($R_{L}$) to the top of a conical conductor. Then we 
move down to its tip at $\vartheta=\vartheta_{0}$ close to the north 
pole of the membrane, and finally down the horizon in the poloidal 
direction to the starting point. (The conical surface is chosen 
instead of an infinitely thin wire in order to avoid divergent 
integrals; it replaces a wire of finite thickness.)

The ideally conducting disk gives no contribution to the EMF. Formally 
this comes about as follows: The 4-velocity field of the disk $u$ 
(whose parts move with the FIDOs) is $e_{0}$ (see (\ref{strau:5})) and 
thus $-i_{u}\,F =E=0$, for an ideal conductor. Using (\ref{strau:7}), this 
implies $\mathcal{E}=0$.

The contribution $V_{L}$ to the line integral $\oint 
\alpha\,\mathcal{E}$ from the load resistance far from the BH is what 
we all know from electrodynamics
\begin{equation}
        V_{L} = I\,R_{L},
        \label{strau:6.1}
\end{equation}
where $I$ is the current in the wire.

There remains the contribution $V_{H}$ to the EMF along the stretched 
horizon
\begin{equation}
        V_{H} = \int_{\mathcal{C}_{H}}\alpha\,\mathcal{E}\stackrel{(\textrm{Ohm})}{=}
        \int_{\mathcal{C}_{H}}  R_{H}\boldsymbol{*}\mathcal{J}_{H}. 
        \label{strau:6.2}
\end{equation}
The current flows along the northern hemisphere of the horizon as a 
surface current in the \emph{poloidal} direction, with surface current 
density
\begin{equation}
        \vec{\mathcal{J}}_{H} = 
        \frac{I}{2\pi\,\tilde{\omega}_{H}}\,\vec{e}_{\vartheta}.
        \label{strau:6.3}
\end{equation}

In order to prove that the surface current $\vec{\mathcal{J}}_{H}$ has 
no toroidal component, we apply (\ref{strau:5.2}) to a toroidal path 
$\{\vartheta =\textrm{const}\}$ on the stretched horizon. Along this 
$i_{\beta}\,\mathcal{B}= -\mathbf{d}\Psi/2\pi =0$, and thus
\begin{displaymath}
        \oint \alpha\,\mathcal{E} =0 \quad \Longrightarrow  \quad
        i_{\partial_{\varphi}}\,\mathcal{E}_{H}=0:\ \vec{E}_{H}\cdot 
        \vec{e}_{\varphi}=0.
\end{displaymath}
Ohm's law in (\ref{strau:69}) then implies $\vec{\mathcal{J}}_{H} \cdot 
\vec{e}_{\varphi} = 0$.

Using (\ref{strau:6.3}) in (\ref{strau:6.2}) gives
\begin{displaymath}
        V_{H} = I\int_{\vartheta_{0}}^{\frac{\pi}{2}}R_{H}\,
        \frac{\rho_{H}}{2\pi\,\tilde{\omega}_{H}}\,d\vartheta
\end{displaymath}
or
\begin{equation}
        V_{H} = I\,R_{HT},
        \label{strau:6.4}
\end{equation}
where
\begin{displaymath}
        R_{HT} = \int_{\vartheta_{0}}^{\frac{\pi}{2}}R_{H}\,
        \frac{\rho_{H}}{2\pi\,\tilde{\omega}_{H}}\,d\vartheta
\end{displaymath}
is the total resistance of the horizon (see (\ref{strau:5.5})). The EMF 
around our closed path is thus
\begin{equation}
        V = V_{H} + V_{L} = I(R_{HT} + R_{L}).
        \label{strau:6.5}
\end{equation}

This voltage is also equal to the line integral on the right hand side 
in (\ref{strau:5.2}),
\begin{equation}
        V = \oint_{\mathcal{C}}i_{\beta}\,\mathcal{B}.
        \label{strau:6.6}
\end{equation}
This receives only contributions from the horizon and the disk (the 
contribution of the latter was overlooked in Ref. [1]). Using 
$\beta=-\omega\,\partial_{\varphi},$ 
$\mathcal{B}_{\perp}=B_{r}\,\vartheta^{\vartheta} \wedge 
\vartheta^{\varphi}$, $\vartheta^{\vartheta}=\rho\,d\vartheta,$ $\vartheta^{\varphi}
=\tilde{\omega}\,d\varphi$, the horizon gives
\begin{equation}
        \int_{\mathcal{C}_{H}}i_{\beta}\,\mathcal{B} = \Omega_{H}
        \int_{\vartheta_{0}}^{\frac{\pi}{2}}B_{\perp}\,\tilde{\omega}_{H}\,\rho_{H}
        \,d\vartheta.
        \label{strau:6.7}
\end{equation}
A similar contribution is obtained along the disk ($\mathcal{C}_{D}$) 
and the total voltage is given by
\begin{equation}
        V = \Omega_{H}\int_{\vartheta_{0}}^{\frac{\pi}{2}}
        B_{\perp}\,\tilde{\omega}_{H}\,\rho_{H}\,d\vartheta
         - \int_{r_{H}}^{r_{D}}B_{\perp}\,\omega\,
        \tilde{\omega}\,\frac{\rho}{\sqrt{\Delta}}\,dr
        \label{strau:6.8}
\end{equation}
($r_{D}$ is the edge of the disk). This ``battery'' 
voltage\footnote{The part of the integral (\ref{strau:6.6}) from the disk 
is independent of the connecting path, because the induction law 
gives $\mathbf{d}(\alpha\,\mathcal{E}-i_{\beta}\,\mathcal{B})=0$ and thus 
$\mathbf{d}\,(i_{\beta}\,\mathcal{B})=0$ inside the disk.}, and the total 
horizon and load resistances $R_{HT}$ and $R_{L}$ determine the 
current $I$ according to (\ref{strau:6.5}).

If the magnetic field is axisymmetric, we can use (\ref{strau:5.6}) to write 
the integrand in (\ref{strau:6.6}) as follows
\begin{equation}
        i_{\beta}\,\mathcal{B}= \frac{\omega}{2\,\pi}\,\mathbf{d}\Psi.
        \label{strau:6.9}
\end{equation}
The voltage is then
\begin{equation}
        V = \frac{\Omega_{H}}{2\,\pi}\,\Psi(\textrm{eq})\left[1 +       
        \int_{horizon}^{edge}\left(\frac{\omega}{\Omega_{H}}\right)\mathbf{d}
        \left(\frac{\Psi}{\Psi(\textrm{eq})}\right)\right],
        \label{strau:6.10}
\end{equation}
where $\Psi(\textrm{eq})$ denotes the value of the flux function at 
the equator of the horizon. The two pieces in the square bracket are 
comparable in magnitude.

We shall see in the next section in detail how the power, dissipated 
as ohmic losses
\begin{equation}
        P_{L} = I^{2}\,R_{L}
        \label{strau:6.11}
\end{equation}
in the load, is extracted from the hole, but this is clearly at the 
cost of the mass of the BH:
\begin{equation}
        I^{2}R_{L} = -\dot{M}.
        \label{strau:6.12}
\end{equation}

Let us note that
\begin{equation}
        P_{L} = V^{2}\,\frac{R_{L}}{(R_{HT} + R_{L})^{2}}.
        \label{strau:6.13}
\end{equation}
This becomes maximal for
\begin{equation}
        R_{HT} = R_{L}\qquad (\textrm{impedance matching})
        \label{strau:6.14}
\end{equation}
with
\begin{equation}
        P_{L}^{max} = \frac{V^{2}}{4\,R_{HT}}.
        \label{strau:6.15}
\end{equation}
This maximal extraction rate was stated in (\ref{strau:5.4}).

\begin{figure}[htbp]
        \centerline{\includegraphics[width=15cm , height=10cm, angle=-1 ]{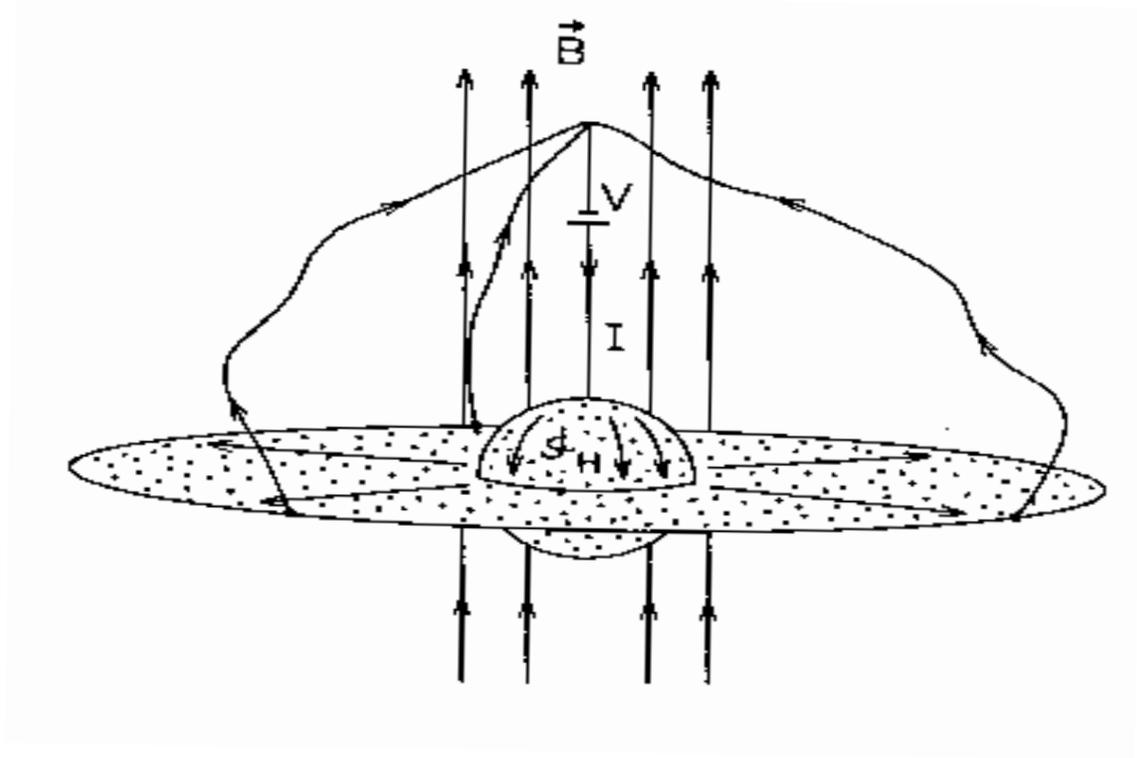}}
        \caption{Black hole playing the role of the rotator of an electric 
        motor.}
        \protect\label{strau:Fig.6.2}
\end{figure}

Clearly, we can also reverse our gedanken experiment. By applying a 
voltage, we can use the BH as the rotator of an electric motor (see 
Fig. \ref{strau:Fig.6.2}). You see that some of the physics of BHs is indeed very 
similar to that of ordinary electric generators and electric motors.

\section{Conservation Laws, Increase of Entropy of a BH}

The general results which will be obtained in this section will 
enable us to develop a more profound analysis of the idealized current 
generator discussed above.

Contraction of the energy-momentum tensor $T^{\mu \nu}$ with the two 
Killing fields $\partial_{t},\ \partial_{\varphi}$ gives two conserved 
vector fields, which express the conservation laws of energy and 
angular momentum in the z-direction. We want to formulate these in the 
3+1 splitted form.

To this end, we consider as a preparation a 4-dimensional equation of 
the type
\begin{equation}
        \nabla \cdot J = Q 
        \label{strau:7.1}
\end{equation}
for a vector field $J$ with source term $Q$. The Hodge dual of the 
1-form $J^{\flat}$ has the same decomposition as $\mathcal{S}$ in 
(\ref{strau:22}):
\begin{equation}
                *J^{\flat} =\rho + (i_{\beta}\,\rho - \alpha\,\mathcal{J}) \wedge dt.
        \label{strau:7.2}
\end{equation}
If $J=J^{\mu}e_{\mu}$, then $\rho$ is the 3-form belonging to 
$J^{0}$, and $\mathcal{J}$ is the 2-form corresponding to 
$\vec{j}=J^{k}e_{k}$. Eq. (\ref{strau:7.1}) is equivalent to  
$d\,\left(*J^{\flat} \right)=Q\,\textrm{vol}_{4}$ or
\begin{equation}
        dt \wedge [(\partial_{t}-L_{\beta})\rho + 
        \mathbf{d}(\alpha\,\mathcal{J})] = Q\,\textrm{vol}_{4}.
        \label{strau:7.3}
\end{equation}
Since $dt=\theta^{0}/\alpha$, the 3+1 split is
\begin{equation}
        (\partial_{t}-L_{\beta})\,\rho + \mathbf{d}(\alpha\,\mathcal{J}) = 
        \alpha\,Q\,\textrm{vol}_{3}.
        \label{strau:7.4}
\end{equation}
If $\textrm{div}\,\vec{\beta}=0$ (as for the Kerr solution), this is 
equivalent to ($\rho =: \hat{\rho}\,\textrm{vol}_{3}$)
\begin{equation}
        (\partial_{t}-L_{\vec{\beta}})\,\hat{\rho} + \vec{\nabla}\cdot 
        \left(\alpha\,\vec{j}\right) = \alpha\,Q.
        \label{strau:7.5}
\end{equation}
The integral form
\begin{equation}
        \frac{d}{dt}\int_{\mathcal{V}}\rho = -\int_{\partial 
        \mathcal{V}}(\alpha\,\mathcal{J}-i_{\beta}\,\rho) + 
        \int_{\mathcal{V}}\alpha\,Q\,\textrm{vol}_{3}
        \label{strau:7.6}
\end{equation}
generalizes the conservation law (\ref{strau:28}). We also write (\ref{strau:7.6}) 
in vector analytic form
\begin{equation}
        \frac{d}{dt}\int_{\mathcal{V}}\hat{\rho}\,dV = -\int_{\partial 
        \mathcal{V}}(\alpha\,\vec{j}- \hat{\rho}\,\vec{\beta}) \cdot 
        d\vec{A} + \int_{\mathcal{V}}\alpha\,Q\,dV.
        \label{strau:7.7}
\end{equation}

Let us now apply this to the vector fields $k_{\mu}T^{\mu\nu}$ und
 $m_{\mu}T^{\mu\nu}$ $(k=\partial_{t}, \  
m=\partial_{\varphi})$. The corresponding $\hat{\rho}$ and $\vec{j}$ 
are denoted as follows
\begin{eqnarray}
        -k\cdot T & \longleftrightarrow & (\varepsilon_{E_{\infty}},\,
        \vec{S}_{E_{\infty}}), \nonumber \\
        m\cdot T & \longleftrightarrow & (\varepsilon_{L_{z}},\, 
        \vec{S}_{L_{z}}).
        \label{strau:7.8}
\end{eqnarray}
Eq. (\ref{strau:7.5}) gives
\begin{eqnarray}
        (\partial_{t}-L_{\vec{\beta}})\,\varepsilon_{E_{\infty}} + \vec{\nabla}\cdot 
        (\alpha\,\vec{S}_{E_{\infty}}) & = & 0 \qquad 
        \textrm{(Poynting theorem)},
        \label{strau:7.9}  \\
        (\partial_{t}-L_{\vec{\beta}})\,\varepsilon_{L_{z}} + \vec{\nabla}\cdot 
        (\alpha\,\vec{S}_{L_{z}}) & = & 0\quad \textrm{(angular momentum 
        conservation).}
        \label{strau:7.10}
\end{eqnarray}
The corresponding integral formulas are
\begin{eqnarray}
        \frac{d}{dt}\int_{\mathcal{V}}\varepsilon_{E_{\infty}}\,dV & = & 
         -\int_{\partial \mathcal{V}}\left(\alpha\,\vec{S}_{E_{\infty}}- 
        \varepsilon_{E_{\infty}}\,\vec{\beta}\right) \cdot d\vec{A}, 
        \label{strau:7.11}  \\
        \frac{d}{dt}\int_{\mathcal{V}}\varepsilon_{L_{z}}\,dV& = & 
        -\int_{\partial \mathcal{V}}\left(\alpha\,\vec{S}_{L_{z}}- 
        \varepsilon_{L_{z}}\,\vec{\beta}\right) \cdot d\vec{A}. 
        \label{strau:7.12}
\end{eqnarray}

Now we make use of (\ref{strau:5}), i.e., $\partial_{t}=\alpha\,e_{0}-\omega\,
\partial_{\varphi}= \alpha\,e_{0} - 
\omega\,\tilde{\omega}\,\vec{e}_{\varphi}$, giving us
\begin{displaymath}
        -k \cdot T = -\alpha\,e_{0} \cdot T + \omega\,m \cdot T.
\end{displaymath}
Here we use the following FIDO decomposition of $T$:
\begin{equation}
        T = \varepsilon\,e_{0}\otimes e_{0} + e_{0} \otimes \vec{S} + \vec{S}
        \otimes e_{0} + \stackrel{\leftrightarrow}{\mathbf{T}},
        \label{strau:7.13}
\end{equation}
and obtain the relations
\begin{eqnarray}
        \varepsilon_{E_{\infty}}& = & \alpha\,\varepsilon + 
        \omega\,\varepsilon_{L_{z}}, \qquad \textrm{(energy density ``at 
        infinity'')},
        \label{strau:7.14}  \\
        \vec{S}_{E_{\infty}} & = & \alpha\,\vec{S} + 
        \omega\,\vec{S}_{L_{z}},\quad 
        \textrm{(energy current density ``at infinity'')}.
        \label{strau:7.15}
\end{eqnarray}

\begin{center}
        \large{\textbf{Application to a Kerr BH}}
\end{center}

The global conservation laws (\ref{strau:7.11}) and (\ref{strau:7.12}) are now 
applied to a Kerr BH. Its mass plays the role of the ``energy at 
infinity'' and thus, the energy conservation (\ref{strau:7.11}) with 
(\ref{strau:7.15}), gives 
\begin{equation}
        \frac{dM}{dt} = 
        -\int_{\mathcal{H}^{s}}\alpha_{H}\,\vec{S}_{E_{\infty}}
        \cdot \vec{n}\,dA = -\int_{\mathcal{H}^{s}}
        \left(\alpha^{2}_{H}\,\vec{S}+\alpha_{H}\,\Omega_{H}\,\vec{S}_{L_{z}}\right)
        \cdot \vec{n}\,dA.
        \label{strau:7.16}
\end{equation}
Similarly, the change of the angular momentum of the BH is by 
(\ref{strau:7.12})
\begin{equation}
        \frac{dJ}{dt} = -\int_{\mathcal{H}^{s}}\alpha_{H}\,\vec{S}_{L_{z}}
        \cdot \vec{n}\,dA; \qquad \vec{S}_{L_{z}} = 
        \vec{\partial}_{\varphi}\cdot \stackrel{\leftrightarrow}{\mathbf{T}}.
        \label{strau:7.17}
\end{equation}

Now, we consider the entropy increase of the BH. The first law \cite{7} 
tells us that
\begin{equation}
        T_{H}\,\frac{dS_{H}}{dt} = \frac{dM}{dt} - \Omega_{H}\,\frac{dJ}{dt}.
        \label{strau:7.18}
\end{equation}
On the right hand side we insert (\ref{strau:7.16}) and (\ref{strau:7.17}), and 
use also (\ref{strau:7.15}), giving us the generally valid formula
\begin{equation}
        T_{H}\,\frac{dS_{H}}{dt} = -\int_{\mathcal{H}^{s}}
        \alpha^{2}_{H}\,\vec{S}\cdot \vec{n}\,dA.
        \label{strau:7.19}
\end{equation}

Until now we have not specified the matter content. For electrodynamics 
we have $\vec{S}= \frac{1}{4\pi}\,\vec{E}\times \vec{B}$, and thus at 
the horizon $\alpha^{2}_{H}\,\vec{S}= \frac{1}{4\pi}\,\vec{E}_{H}\times \vec{B}_{H}$
(see (\ref{strau:77})). In this case we obtain with the laws of Amp\`{e}re 
and Ohm
\begin{equation}
        T_{H}\,\frac{dS_{H}}{dt} =  -\frac{1}{4\pi}\int_{\mathcal{H}^{s}}
        (\vec{E}_{H}\times \vec{B}_{H})\cdot \vec{n}\,dA = \int_{\mathcal{H}^{s}}
        \vec{\mathcal{J}}_{H}\cdot \vec{E}_{H}\,dA = \int_{\mathcal{H}^{s}}
        R_{H}\,\vec{\mathcal{J}}^{2}_{H}\,dA.
        \label{strau:7.20}
\end{equation}
All of these familiarly looking expressions for the rate of entropy 
increase are important and useful.

Let us also evaluate (\ref{strau:7.17}) in a similar manner. First, we have
\begin{equation}
        \vec{S}_{L_{z}} = \frac{1}{4\pi}\,\vec{\partial}_{\varphi}\cdot 
        \left[-\left(\vec{E}\otimes \vec{E} + \vec{B}\otimes \vec{B}\right) 
        +\frac{1}{2}\left(\vec{E}^{2} + 
        \vec{B}^{2}\right)\stackrel{\leftrightarrow}{\mathbf{g}}\right].
        \label{strau:7.21}
\end{equation}
Clearly, only the first term contributes to the integrand in 
(\ref{strau:7.17}). Using this time the laws of Gauss and Amp\`{e}re, we 
find
\begin{equation}
        \frac{dJ}{dt} = \int_{\mathcal{H}^{s}}\left(\sigma_{H}\,\vec{E}_{H} 
        + \vec{\mathcal{J}}_{H} \times \vec{B}_{n}\right)\cdot 
        \vec{\partial}_{\varphi}\,dA. 
        \label{strau:7.22}
\end{equation}
Note that the first term is absent if $\vec{E}_{H}$ has no toroidal 
component.

Finally, we use (\ref{strau:7.18}), together with (\ref{strau:7.20}) and 
(\ref{strau:7.22}), to obtain the following formula for the change of the 
mass of the BH 
\begin{equation}
        \frac{dM}{dt} = 
        \int_{\mathcal{H}^{s}}\left[\vec{\mathcal{J}}_{H}\cdot \vec{E}_{H}
        -\vec{\beta}_{H}\cdot \left(\sigma_{H}\,\vec{E}_{H} 
        + \vec{\mathcal{J}}_{H} \times \vec{B}_{n}\right)\right]dA.
        \label{strau:7.23}
\end{equation}

\begin{center}
        \large{\textbf{Application to the idealized current generator}}
\end{center}

It is instructive to use these general results for a more detailed 
analysis of the current generator, discussed in the last section.

We begin by computing the ohmic heating rate of the current flowing 
through the northern hemisphere (n.H.) of the BH:
\begin{eqnarray}
        \int_{n.H.}\vec{E}_{H}\cdot \vec{\mathcal{J}}_{H}\,dA & = & 
        \int_{\vartheta_{0}}^{\frac{\pi}{2}}E_{H\hat{\vartheta}}\,\mathcal{J}_{H\hat{\vartheta}}\,
        2\pi\,\tilde{\omega}\,\rho_{H}\,d\vartheta  \nonumber  \\
         & \stackrel{(\ref{strau:6.3})}{=} & 
         I\int_{\vartheta_{0}}^{\frac{\pi}{2}}E_{H\hat{\vartheta}}\,\rho_{H}\,d\vartheta
        = I\,V_{H} \stackrel{(\ref{strau:6.4})}{=} I^{2}\,R_{HT}.
        \label{strau:7.24}
\end{eqnarray}
According to the general result (\ref{strau:7.20}) this rate is equal to $T_{H}\,dS_{H}/dt$. 
Thus,
\begin{equation}
        T_{H}\,\frac{dS_{H}}{dt} = I^{2}\,R_{HT}.
        \label{strau:7.25}
\end{equation}

\begin{figure}[htbp]
        \centerline{\includegraphics[width=14cm , height=8cm ]{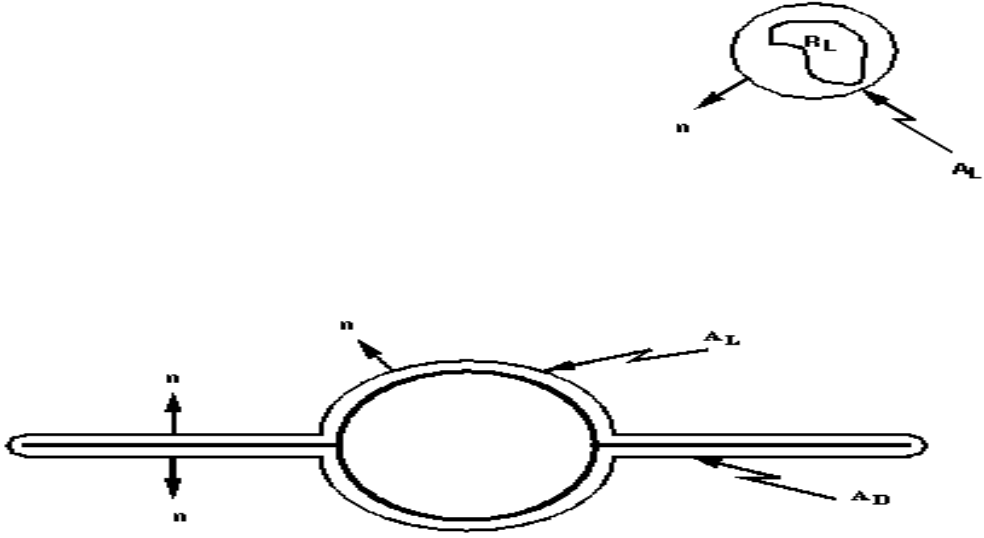}}
        \caption{The boundary $\partial V=\mathcal{A}_{H}\cup 
        \mathcal{A}_{D}\cup \mathcal{A}_{L}$ for (\ref{strau:7.27})}
        \protect\label{strau:Fig.7.1}
\end{figure}

In order to trace the details of the energy flow, we apply the 
generalized Poynting theorem (\ref{strau:7.11}). For a stationary situation 
this reduces to
\begin{equation}
        \int_{\partial \mathcal{V}}\left(\alpha\,\vec{S}_{E_{\infty}}- 
        \varepsilon_{E_{\infty}}\,\vec{\beta}\right) \cdot d\vec{A} = 0.
        \label{strau:7.26}
\end{equation}
We choose the volume $\mathcal{V}$ such that the boundary 
$\partial\,V$ consists of a horizon part $\mathcal{A}_{H}$ 
($\mathcal{H}^{s}$ minus the inner edge of the disk), the disk 
$\mathcal{A}_{D}$, and a surface $\mathcal{A}_{L}$ enclosing the load's 
resistor (see Fig. \ref{strau:Fig.7.1}). The second term in (\ref{strau:7.26}) does not 
contribute for $\mathcal{A}_{H}$ and $\mathcal{A}_{D}$, and can be 
ignored for the load, since this is assumed to be located far from 
the horizon. Using also the relation (\ref{strau:7.15}) we have then 
\begin{equation}
        \int_{\partial\mathcal{V}}\alpha\,\vec{S}_{E_{\infty}}\cdot d\vec{A}=\int_{\partial\mathcal{V}}
        (\alpha^{2}\vec{S}+\alpha\,\omega\,\vec{S}_{L_{z}})
        \cdot d\vec{A}= 0.
        \label{strau:7.27}
\end{equation}

The contribution from the horizon to the first term on the right is
\begin{equation}
        -\int_{\mathcal{A}_{H}}\alpha^{2}\vec{S}\cdot \vec{n}\,dA 
        \stackrel{(\ref{strau:7.19})}{=} T_{H}\,\frac{dS_{H}}{dt} 
        \stackrel{(\ref{strau:7.25})}{=} I^{2}R_{HT}, 
        \label{strau:7.28}
\end{equation} 
and the second term gives
\begin{equation}
        -\Omega_{H}\int_{\mathcal{A}_{H}}\alpha_{H}\,\vec{S}_{L_{z}}
        \cdot \vec{n}\,dA.
        \label{strau:7.29}
\end{equation}
At first sight one might think that this is just 
$\Omega_{H}\,\frac{dJ}{dt}$ (see (\ref{strau:7.17})). This is, however, not 
correct, because there is an additional inflow of angular momentum 
through the disk. Physically, this is clear: The external magnetic 
field exerts -- through the Lorentz force on the surface current 
$\vec{\mathcal{J}}_{D}$ of the disk -- a torque on the disk, and thus 
on the BH to which the disk is locked. As in the derivation of 
(\ref{strau:7.22}) we find for (\ref{strau:7.29})
\begin{eqnarray*} 
        -\Omega_{H}\int_{\mathcal{A}_{H}}\alpha_{H}\,\vec{S}_{L_{z}}
        \cdot \vec{n}\,dA &=&
        \Omega_{H}\int_{\mathcal{A}_{H}}\tilde{\omega}_{H}(\vec{\mathcal{J}}_{H}\times \vec{B}_{n})
        \cdot \vec{e}_{\varphi} \,dA \\  =
        -\Omega_{H}\int_{\vartheta_{0}}^{\frac{\pi}{2}}\mathcal{J}_{H\hat{\vartheta}}\, 
        B_{\perp}\,\tilde{\omega}_{H}\, 2\pi \,\tilde{\omega}_{H} 
        \,\rho_{H}\,d\vartheta   &
    \stackrel{(\ref{strau:6.3})}{=}& -I\,\Omega_{H}\int_{\vartheta_{0}}^{\frac{\pi}{2}}
    B_{\perp}\, \tilde{\omega}_{H}\,\rho_{H}\,d\vartheta,
\end{eqnarray*}
where use has been made of the fact, that $\vec{E}_{H}$ has no toroidal 
component. Together with (\ref{strau:6.7}), we obtain for this horizon 
contribution
\begin{equation}
        -\Omega_{H}\int_{\mathcal{A}_{H}}\alpha_{H}\,\vec{S}_{L_{z}}
        \cdot \vec{n}\,dA = -I\int_{\mathcal{C}_{H}}i_{\beta}\,\mathcal{B}.
        \label{strau:7.30}
\end{equation}
Note that the integral on the right is not the total voltage 
(\ref{strau:6.6}). However, we shall see shortly that the disk contributes 
the remaining part of $-I\,V$.

Only the last term in (\ref{strau:7.27}) gets contributions from the disk 
(since $\vec{E}=0$ in the disk). Using the expression (\ref{strau:7.21}) 
for $\vec{S}_{L_{z}}$, this becomes
\begin{equation}
        -\int_{\mathcal{A}_{D}}\alpha\,\omega \,\vec{S}_{L_{z}} \cdot \vec{n}\,dA =
        \int_{\mathcal{C}_{D}}\alpha\,\omega 
        \,\frac{B_{n}}{4\pi}\,\vec{\partial}_{\varphi}\cdot 
        \vec{B}\,2\pi\tilde{\omega}\,dl.
        \label{strau:7.31}
\end{equation}
Between the parallel component $\vec{B}_{\parallel}$ and the surface 
current $\vec{\mathcal{J}}_{D}$ of the disk we have Amp\`{e}re's 
relation of ordinary electrodynamics:
\begin{equation}
        \alpha\,\vec{B}_{\parallel} = 4\pi \,\vec{\mathcal{J}}_{D}\times 
        \vec{n} = 4\pi\,\frac{I}{2\pi\,\tilde{\omega}}\,\vec{e}_{r}\times 
        \vec{n}.
        \label{strau:7.32}
\end{equation}
Making use of this, (\ref{strau:7.31}) becomes
\begin{displaymath}
         -\int_{\mathcal{A}_{D}}\alpha\,\omega \,\vec{S}_{L_{z}} \cdot \vec{n}\,dA  = 
         -I\int_{\mathcal{C}_{D}}\vec{\beta}\cdot\left(\vec{e}_{r}\times 
         \vec{B}_{n}\right)\,dl = 
         I\int_{\mathcal{C}_{D}}\vec{\beta}\cdot\left(\vec{\beta}\times 
         \vec{B}_{n}\right)\cdot d\vec{l}   
\end{displaymath}
or
\begin{equation}
         -\int_{\mathcal{A}_{D}}\alpha\,\omega \,\vec{S}_{L_{z}} \cdot 
         \vec{n}\,dA =  -\int_{\mathcal{C}_{D}}i_{\beta}\,\mathcal{B},
        \label{strau:7.33}
\end{equation}
as already announced.

Finally, the contribution to (\ref{strau:7.27}) is what we are used to:
\begin{equation}
        \int_{\mathcal{A}_{L}}\alpha\,\vec{S}_{E_{\infty}} \cdot d\vec{A} =
        \int_{\mathcal{A}_{L}}\frac{1}{4\pi}\,(\vec{E}\times \vec{B})\,d\vec{A} = 
        I^{2}R_{L}.
        \label{strau:7.34}
\end{equation}

All together, the Poynting theorem (\ref{strau:7.27}) gives
\begin{equation}
        \underbrace{I^{2}R_{TH} 
        -I\int_{\mathcal{C}_{H}}i_{\beta}\,\mathcal{B}}_{\textrm{horizon}} -    
        \underbrace{I\int_{\mathcal{C}_{D}}i_{\beta}\,\mathcal{B}}_{\textrm{disk}}
         = \underbrace{-I^{2}R_{L}}_{\textrm{load}}.
        \label{strau:7.35}
\end{equation}

>From the derivation it is clear how to interpret this result. First, 
we note that the left hand side is the energy (measured at infinity) 
which flows per unit time into the BH, and thus is equal to $dM/dt$. 
According to (\ref{strau:7.25}) the first term is $T_{H}\,dS_{H}/dt$ (ohmic 
dissipation). The second term in (\ref{strau:7.35}) is that part of 
$\Omega_{H}\,dJ/dt$ which is due to the torque acting on the surface 
current density $\vec{\mathcal{J}}_{H}$ (see (\ref{strau:7.17}) and 
(\ref{strau:7.22})). The third term resulted from the disk and gives an 
additional contribution to the change of the rotational energy, which 
flows through the disk into the BH. For clarification we note that for 
a closed surface $\mathcal{A}$ surrounding the disk we have
\begin{displaymath}
                \oint_{\mathcal{A}} \alpha\,\vec{S}_{E_{\infty}}\cdot d\vec{A} =0= \oint_{\mathcal{A}} 
        \alpha\,\omega \,\vec{S}_{L_{z}}\cdot d\vec{A},
\end{displaymath}
since $\vec{S}$ vanishes in the disk. Therefore, the rotational 
energy which flows through the inner edge of the disk into the BH is 
(see (\ref{strau:7.17}))
\begin{displaymath}
        \Omega_{H}\int_{\gets \mathbf{|}}\alpha\,\vec{S}_{L_{z}}\cdot d\vec{A} =
        -\int_{\mathcal{A}_{D}}\alpha\,\omega \,\vec{S}_{L_{z}} \cdot \vec{n}\,dA 
        \stackrel{(\ref{strau:7.33})}{=}-\int_{\mathcal{C}_{D}}i_{\beta}\,\mathcal{B}. 
\end{displaymath}

The total change of the rotational energy of the BH thus is
\begin{eqnarray}
        \Omega_{H}\,\frac{dJ}{dt} & = & -I\oint_{\mathcal{C}}i_{\beta}\,\mathcal{B} 
        \stackrel{(\ref{strau:6.6})}{=} -I\,V \nonumber \\
         & \stackrel{(\ref{strau:6.5})}{=}& -I^{2}(R_{HT}+R_{L}).
        \label{strau:7.36}
\end{eqnarray}

The presence of the disk made the discussion a bit complicated, but 
it is nice to see how the various pieces combine. Schematically, we 
have
\begin{equation}
        \begin{array}{ccccc}
                \frac{dM}{dt} & = & T_{H}\,\frac{dS_{H}}{dt} & + & 
                \Omega_{H}\,\frac{dJ}{dt}  \\
                \parallel & & \parallel  &  & \parallel  \\
                -I^{2}R_{L} &  & I^{2}R_{HT} &  & -I^{2}(R_{HT}+R_{L}).
        \end{array}
        \label{strau:box}
\end{equation}
The energy (at $\infty$) $dM/dt$ flows -- partly through the disk -- 
down the hole and is, of course, negative because the spin down 
overcomes the ohmic dissipation. This part of the energy balance 
results, therefore, in an outflowing energy which is carried without 
loss by the electromagnetic fields to the load resistors, where it is 
dissipated at the rate $I^{2}R_{L}$. Netto, we obtain, of course,
\begin{equation}
        -\frac{dM}{dt} = I^{2}R_{L}.
        \label{strau:7.38}
\end{equation}

\section{Blandford-Znajek Process}

Let us return to Fig. \ref{strau:Fig.5.4}, in which a plausible magnetic 
field structure around a supermassive BH in the center of an active 
galaxy is sketched. Rotation and turbulence in an accretion disk can 
generate magnetic fields of the order 1 tesla, as we estimated at the 
end of section 5.

Close to the BH one expects a force-free electron-positron plasma, 
which comes about as follows. Imagine first that there are no charged 
particles in the neighborhood of the hole. In this case unipolar 
induction generates a quadrupole-like electric field similar to that 
we found in section 3 (see Fig. \ref{strau:Fig.3.1}). Close to the BH the 
magnitude of this electric field is $E \sim B\,a/M \sim 3\times 
10^{6}$ (Volt/cm)$\,(a/M)$. Between the horizon and a few 
gravitational radii along the magnetic field lines this gives rise to 
a voltage $V \sim E\,r_{H} \sim B\,a \sim 10^{20}$ Volt for a BH with 
mass $M \sim 10^{9}\,M_{\odot}$ (see section 5). In this enormous 
potential stray electrons from the disk or interstellar space will be 
accelerated along magnetic field lines to ultrarelativistic energies. 
Inverse Compton scattering with soft photons from the accretion disk 
leads to $\gamma$-quanta which in turn annihilate with soft photons 
from the disk into electron-positron pairs. These will again be 
accelerated and by repetition an electron-positron plasma is 
generated which can become dense enough to annihilate the component 
of $\vec{E}$ along $\vec{B}$. The electric field is then nearly 
orthogonal to $\vec{B}$, up to a sufficiently large component which 
produces occasional electron-positron sparks in order to fill the 
magnetosphere with plasma. (Similar processes are important in the 
magnetospheres of pulsars. It is likely that all active pulsars have 
electron-positron winds.)

\begin{figure}[htbp]
        \centerline{\includegraphics[width=8cm , height=6cm ]{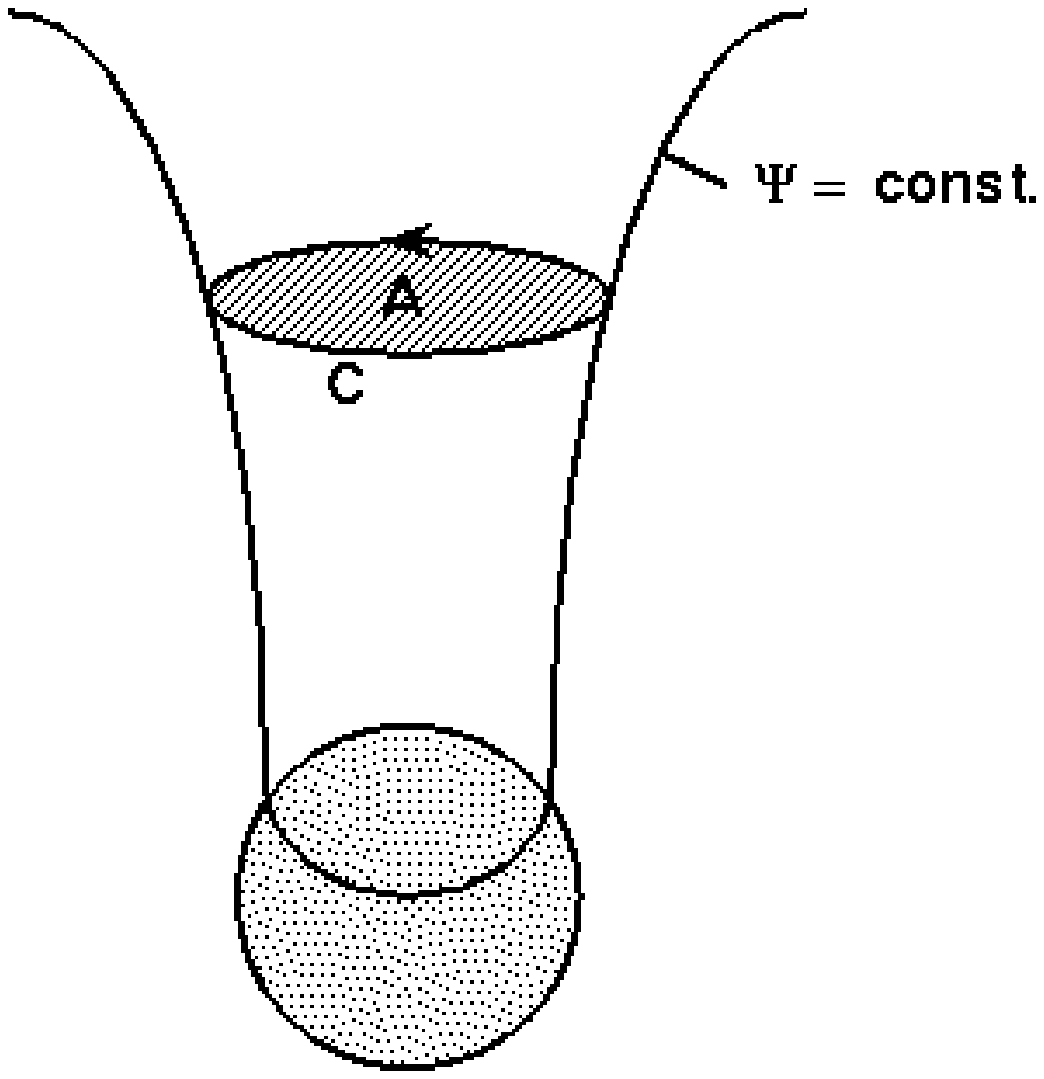}}
        \caption{Path for (\ref{strau:8.2}) and (\ref{strau:8.3}).}
        \protect\label{strau:Fig.8.1}
\end{figure}

Fields with $\vec{E} \cdot \vec{B}=0$ are called \emph{degenerate}. We 
assume in what follows that in the neighborhood of the BH, where the 
$\vec{B}$-field is strong, the electromagnetic field is 
\emph{force-free}, which means that the ideal MHD condition
\begin{equation}
        \rho_{e}\,\vec{E} + \vec{j} \times \vec{B} = 0
        \label{strau:8.1}
\end{equation}
is satisfied. Clearly, in a force-free plasma $\vec{E}$ is 
perpendicular to $\vec{B}$. Furthermore, $\vec{E}$ has no toroidal 
component for an axisymmetric stationary situation. This is an 
immediate consequence of the induction law: Applying (\ref{strau:25}) for a 
stationary and axisymmetric configuration to the path $\mathcal{C}$ in 
Fig \ref{strau:Fig.8.1}, we obtain
\begin{equation}
        \oint_{\mathcal{C}}\alpha\,\mathcal{E} =\oint_{\mathcal{C}}i_{\beta}\,\mathcal{B}      
    \stackrel{(\ref{strau:5.6})}{=}-\frac{1}{2\,\pi}\oint_{\mathcal{C}}\mathbf{d}\Psi = 
    0 \Longrightarrow \vec{E}^{\textrm{tor}}=0.
        \label{strau:8.2}
\end{equation}
Similarly, Amp\`{e}re's law in integral form (\ref{strau:27}) reduces to
\begin{equation}
        \oint_{\mathcal{C}}\alpha\,\mathcal{H} = 4\,\pi 
        \oint_{\mathcal{A}}\alpha\,\mathcal{J} = 4\,\pi\,I,
        \label{strau:8.3}
\end{equation}
where $I$ is the total upward current through a surface $\mathcal{A}$ 
bounded by $\mathcal{C}$. This gives for the toroidal component of the 
magnetic field
\begin{equation}
        \vec{B}^{\textrm{tor}} = 
        \frac{2\,I}{\alpha\,\tilde{\omega}}\,\vec{e}_{\varphi}.
        \label{strau:8.4}
\end{equation}
Using this in (\ref{strau:5.7}) gives
\begin{equation}
        \mathcal{B} = \underbrace{\frac{1}{2\,\pi}\,\mathbf{d}\Psi \wedge 
        \mathbf{d}\varphi}_{\textrm{poloidal}} + 
        \underbrace{\frac{2\,I}{\alpha}\,\boldsymbol{*}\mathbf{d}\varphi}_{\textrm{toroidal}}.
        \label{strau:8.5}
\end{equation}

For later use, we also note the following: The continuity equation 
(\ref{strau:24}) reduces to
\begin{equation}
        \mathbf{d}\,(\alpha\,\mathcal{J}) = 0.
        \label{strau:8.6}
\end{equation}
Moreover, $L_{\partial_{\varphi}}\,(\alpha\,\mathcal{J}) =0$, we 
have $\mathbf{d}\,(i_{\partial_{\varphi}}\,\alpha\,\mathcal{J}) =0$, thus
\begin{equation}
        i_{\partial_{\varphi}}\,(\alpha\,\mathcal{J}) = \frac{1}{2\,\pi} \,\mathbf{d}I
        \label{strau:8.7}
\end{equation}
or
\begin{equation}
        \alpha\,\mathcal{J}^{\textrm{pol}} = \frac{1}{2\,\pi} \,\mathbf{d}I\wedge 
        \mathbf{d}\varphi.
        \label{strau:8.8}
\end{equation}
Clearly, the potential $I$ is the current in (\ref{strau:8.3}).

Because $\mathcal{E}$ is poloidal, we can represent the electric 
field as follows
\begin{equation}
        \mathcal{E} = i_{\vec{v}_{F}}\,\mathcal{B} \qquad (\vec{E} = 
        -\vec{v}_{F} \times \vec{B}),
        \label{strau:8.9}
\end{equation}
where $\vec{v}_{F}$ is toroidal. Let us set
\begin{equation}
        \vec{v}_{F} =: \frac{1}{\alpha}\,(\Omega_{F}-\omega)\,\tilde{\omega}\,\vec{e}
        _{\varphi}.
        \label{strau:8.10}
\end{equation}
For the interpretation of $\Omega_{F}$ note the following: For an 
observer, rotating with angular velocity $\Omega$, the 4-velocity is 
$u=u^{t}\,(\partial_{t}+\Omega\,\partial_{\varphi})$. On the other 
hand, $u=\gamma\,(e_{0}+\vec{v})$, where $\vec{v}$ is the 3-velocity 
relative to a FIDO. Using also 
$\partial_{t}=\alpha\,e_{0}+\vec{\beta}$ we get 
$\Omega\,\vec{\partial}_{\varphi}=\alpha\, \vec{v}-\vec{\beta}$ or
\begin{equation}
        \vec{v} = 
        \frac{1}{\alpha}\,(\Omega-\omega)\,\vec{\partial}_{\varphi} =
        \frac{1}{\alpha}\,(\Omega-\omega)\,\tilde{\omega}\,\vec{e}_{\varphi}.
        \label{strau:8.11}
\end{equation}
This has the same form as (\ref{strau:8.10}) and, therefore, $\Omega_{F}$ is 
the angular velocity of the magnetic field lines.

Next, we show that $\Omega_{F}$ is only a function of $\Psi$. The 
induction law gives
\begin{eqnarray*}
        \mathbf{d}(\alpha\,\mathcal{E}) & = & L_{\beta}\,\mathcal{B} = 
        -L_{\omega\,\partial_{\varphi}}\,\mathcal{B} = 
        -\omega\,L_{\partial_{\varphi}}\,\mathcal{B} -\mathbf{d}\omega \wedge i_{\partial_{\varphi}}
        \,\mathcal{B}  \\
         & = & \frac{1}{2 \,\pi}\,\mathbf{d}\omega \wedge \mathbf{d}\Psi.
\end{eqnarray*}
On the other hand, (\ref{strau:8.9}) and (\ref{strau:8.10}) give
\begin{eqnarray*}
        \mathbf{d}(\alpha\,\mathcal{E}) & = & 
        \mathbf{d}(\alpha\,i_{\vec{v}_{F}}\,\mathcal{B}) = \mathbf{d}\left((\Omega_{F}-\omega)\, 
        i_{\partial_{\varphi}}\,\mathcal{B} \right) = -\mathbf{d}\left( \frac{\Omega_{F}-\omega}
        {2\,\pi}\,\mathbf{d}\Psi \right)\\
         &= &  -\mathbf{d}\left( \frac{\Omega_{F}-\omega}{2\,\pi}\right) \wedge \mathbf{d}\Psi.
\end{eqnarray*}
By comparison, we get $\mathbf{d}\Omega_{F} \wedge \mathbf{d}\Psi =0 \Longrightarrow 
\Omega_{F} = \Omega_{F}(\Psi)$. The calculation above also shows
\begin{equation}
        \alpha \, \mathcal{E} = -\frac{\Omega_{F}-\omega}{2\,\pi}\,\mathbf{d}\Psi,
        \label{strau:8.12}
\end{equation}
i.e., $\vec{E}$ is \emph{perpendicular} to the surfaces $\{\Psi = 
\textrm{const}\}$.

Specializing to the horizon gives
\begin{equation}
        \vec{E}_{H} = 
        -(\Omega_{F}-\Omega_{H})\,\tilde{\omega}\,\vec{e}_{\varphi} \times 
        \vec{B}_{\perp}.
        \label{strau:8.13}
\end{equation}
The representations (\ref{strau:8.5}) of $\mathcal{B}$, and (\ref{strau:8.12}) 
for $\mathcal{E}$ will be important in the final section 9.

\begin{figure}[htbp]
        \centerline{\includegraphics[width=8cm , height=8cm ]{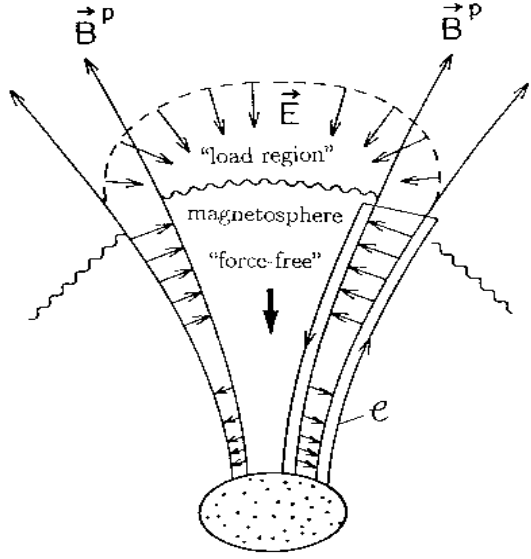}}
        \caption{Magnetosphere of a rotating BH (adapted from [1]).}
        \protect\label{strau:Fig.8.2}
\end{figure}

Now, we proceed as earlier in section 5 and consider again the closed 
path $\mathcal{C}$ in Fig. \ref{strau:Fig.8.2}.

We already had the relations
\begin{equation}
        \triangle V = \frac{1}{2\,\pi}\,\Omega_{H}\,\triangle \Psi,
        \label{strau:8.14}
\end{equation}
and
\begin{equation}
        \triangle R_{H} = R_{H}\,\frac{\triangle 
        \Psi}{4\,\pi^{2}\,\tilde{\omega}^{2}\,B_{\perp}}
        \label{strau:8.15}
\end{equation}
(see (\ref{strau:5.10}) and (\ref{strau:5.13})). As in (\ref{strau:6.4}) we obtain for 
the horizon voltage
\begin{equation}
        \triangle V_{H} = I\,\triangle R_{H}.
        \label{strau:8.16}
\end{equation}
This contribution can alternatively be computed with (\ref{strau:8.12}):
\begin{equation}
        \triangle V_{H} = \int_{\mathcal{C}_{H}}\alpha\,\mathcal{E} =  
        \frac{\Omega_{H}-\Omega_{F}}{2\,\pi}\,\triangle \Psi.
        \label{strau:8.17}
\end{equation}
The total voltage $\triangle V$ of $\mathcal{C}$ is the sum of 
$\triangle V_{H}$ and the voltage drop $\triangle V_{L}$ in the 
astrophysical load region (see Fig. \ref{strau:Fig.8.2}). For the latter we 
obtain from (\ref{strau:8.12}) (since $\omega \simeq 0$)
\begin{equation}
        \triangle V_{L} = \frac{1}{2 \pi}\,\Omega_{F}\,\triangle \Psi
        \label{strau:8.18}
\end{equation}
and, of course, also
\begin{equation}
        \triangle V_{L} = I \,\triangle R_{L},
        \label{strau:8.19}
\end{equation}
where $\triangle R_{L}$ is the resistance of the load region. From 
this equations, and
\begin{equation}
        \triangle V = \triangle V_{H} + \triangle V_{L},
        \label{strau:8.20}
\end{equation}
we immediately find the relations
\begin{eqnarray}
        \frac{\triangle V_{L}}{\triangle V_{H}} & = & 
        \frac{\Omega_{F}}{\Omega_{H}-\Omega_{F}} = \frac{\triangle R_{L}}{\triangle R_{H}},     
        \label{strau:8.21}  \\
        I & = & \frac{\triangle V}{\triangle R_{H}+\triangle R_{L}} = 
        \frac{1}{2}\,(\Omega_{H}-\Omega_{F})\,\tilde{\omega}^{2}\,B_{\perp}.
        \label{strau:8.22}
\end{eqnarray}

The ohmic dissipation at the horizon is as in (\ref{strau:7.25})
\begin{eqnarray}
        T_{H}\,\frac{d\,(\triangle S_{H})}{dt} & = & I^{2}\,\triangle R_{H} =
        I\, \triangle V_{H}     \nonumber  \\
         & = & \frac{(\Omega_{H}-\Omega_{F})^{2}}{4\,\pi}\,\tilde{\omega}^{2}\,B_{\perp}
         \,\triangle \Psi.
        \label{strau:8.23}
\end{eqnarray}
The power $\triangle P_{L}$ deposited in the load is thus
\begin{equation}
        \triangle P_{L} = I^{2}\,\triangle R_{L} = I\, \triangle V_{L}
        \stackrel{(\ref{strau:8.18}),\,(\ref{strau:8.22})}{=} \frac{\Omega_{F}\,(\Omega_{H}-\Omega_{F})}
        {4\,\pi}\,\tilde{\omega}^{2}\,B_{\perp} \,\triangle \Psi.
        \label{strau:8.24}
\end{equation}
As in (\ref{strau:6.14}), $\triangle P_{L}$ becomes maximal for $\triangle 
R_{H}=\triangle R_{L}$, which is the case for (see (\ref{strau:8.21}))
\begin{equation}
        \Omega_{F} = \frac{1}{2}\,\Omega_{H}.
        \label{strau:8.25}
\end{equation}
Then the estimates (\ref{strau:5.16}) and (\ref{strau:5.18}) hold.

Whether the crucial condition (\ref{strau:8.25}) is approximately satisfied 
in realistic astrophysical scenarios is a difficult problem for model 
builders. Therefore, the question remains open whether the process 
proposed by Blandford and Znajek \cite{9} is important for the energy 
production in active galactic nuclei. It could, however, play a 
significant role in the formation of relativistic jets.

In this connection it should be mentioned that the Blandford-Znajek 
process may play an important role in gamma-ray bursts, as has been 
suggested, for instance by M\'{e}sz\'{a}ros and Rees \cite{10}. One of the 
proposed models involves the toroidal debris from a disrupted neutron 
star orbiting around a BH. This debris may contain a strong magnetic 
field, perhaps amplified by differential rotation, and an axial 
magnetically-dominated wind may be generated along the rotation axis 
which would contain little baryon contamination. Energized by the BH 
via the Blandford-Znajek process, a narrow channel Poynting-dominated 
outflow with little baryon loading could be formed. The latter 
property is crucial for explaining the efficient radiation as gamma 
rays. (For a review of this topic, see for instance \cite{11}.)

\section{Axisymmetric, Stationary Electrodynamics of Force-Free Fields}

In this concluding section we discuss the electrodynamics of 
force-free fields around a BH a bit more systematically. The main goal 
is to derive the general relativistic \emph{Grad-Shafranov equation}, 
whose nonrelativistic limit plays an important role in the 
electrodynamics of pulsars.

For \emph{stationary} fields, Maxwell's equations (\ref{strau:38}) become
\begin{eqnarray}
        \mathbf{d}\mathcal{B}=0, &  &\mathbf{d}(\alpha\,\mathcal{E}) = 
     L_{\beta}\,\mathcal{B}, \label{strau:9.1} \\
        \mathbf{d}\boldsymbol{*}\mathcal{E} = 4\pi \rho, &  & 
        \mathbf{d}(\alpha 
        \boldsymbol{*}\mathcal{B}) = 4 \,\pi\, 
        \alpha\,\mathcal{J}-L_{\beta}\boldsymbol{*}\mathcal{E}  .
        \label{strau:9.2}
\end{eqnarray}
>From now on we assume also \emph{axisymmetry}. Then $L_{\beta}\,\rho = 
\mathbf{d}\,i_{\beta}\,\rho = 0$, and hence the continuity equation 
(\ref{strau:24}) reduces to
\begin{equation}
        \mathbf{d}\,(\alpha\,\mathcal{J})=0.
        \label{strau:9.3}
\end{equation}

I recall the representation (\ref{strau:8.5}) of $\mathcal{B}$ in terms of 
the potential $\Psi$ and $I$:
\begin{equation}
        \mathcal{B} = \frac{1}{2\,\pi}\,\mathbf{d}\Psi \wedge 
        \mathbf{d}\varphi +\frac{2\,I}{\alpha}\boldsymbol{*}\mathbf{d}\varphi.
        \label{strau:9.4}
\end{equation}
$\Psi$ and $I$ are independent of $\varphi$; their physical 
significance has already been discussed: Using the notation in Fig. 
\ref{strau:Fig.8.1} we have
\begin{equation}
        \Psi = \int_{\mathcal{A}}\mathcal{B},\qquad 
        I=\int_{\mathcal{A}}\alpha\,\mathcal{J}.
        \label{strau:9.5}
\end{equation}

At this point we make the simplifying assumption that the ideal MHD 
condition (\ref{strau:8.1}) holds \emph{everywhere}. The fields are thus 
force-free, and we know from section 8 that the $\mathcal{E}$-field 
is poloidal and can be represented as (see (\ref{strau:8.12}))
\begin{equation}
        \alpha \, \mathcal{E} = -\frac{\Omega_{F}-\omega}{2\,\pi}\,\mathbf{d}\Psi,
        \label{strau:9.6}
\end{equation}

The representation (\ref{strau:8.8}) of the poloidal current
\begin{equation}
        \alpha\,\mathcal{J}^{\textrm{pol}} = \frac{1}{2\,\pi} \,\mathbf{d}I\wedge 
        \mathbf{d}\varphi.
        \label{strau:9.7}
\end{equation}
was obtained without assuming the ideal MFD condition. But if this is 
assumed, the toroidal part of $\vec{j}\times \vec{B}$ in (\ref{strau:8.1}) 
has to vanish, which means that $\mathcal{J}^{\textrm{pol}}$ and 
$\mathcal{B}^{\textrm{pol}}$ are proportional to each other. 
Comparison of (\ref{strau:9.4}) and (\ref{strau:9.7}) shows that $\mathbf{d}I$ must then be 
proportional to $\mathbf{d}\Psi$, thus $I$ is a function of $\Psi$ alone:
\begin{equation}
        I = I(\Psi).
        \label{strau:9.8}
\end{equation}
We have shown already in section 8 that $\Omega_{F}$ is also a 
function of $\Psi$,
\begin{equation}
        \Omega_{F} = \Omega_{F}(\Psi).
        \label{strau:9.9}
\end{equation}
According to (\ref{strau:9.4}) and (\ref{strau:9.7}) we now have
\begin{equation}
        \mathcal{J}^{\textrm{pol}} = \frac{1}{\alpha}\,\frac{dI}{d\Psi}\,
        \mathcal{B}^{\textrm{pol}}.
        \label{strau:9.10}
\end{equation}

The total current $\vec{j}$ can now be represented as
\begin{equation}
        \vec{j} = \rho_{e}\,\vec{v}_{F} +  
        \frac{1}{\alpha}\,\frac{dI}{d\Psi}\,\vec{B}.
        \label{strau:9.11}
\end{equation}
In view of (\ref{strau:9.10}), the poloidal part of this equation is 
certainly correct, and the toroidal part on the right hand side is 
chosen such that $\vec{j} \times \vec{B} = \rho_{e}\,\vec{v}_{F}\times 
\vec{B} = -\rho_{e}\,\vec{E}$ (see (\ref{strau:8.9})), which is just the 
ideal MFD condition.

Our goal is now to derive -- for given functions (\ref{strau:9.8}) and 
(\ref{strau:9.9}) -- a partial differential equation for the potential $\Psi$. 
We shall achieve this by computing the toroidal part of the current in 
two independent ways. First, we take the toroidal part of (\ref{strau:9.11}) 
and obtain with $B^{\textrm{tor}} 
=\frac{2\,I}{\alpha\,\tilde{\omega}}$ (see (\ref{strau:9.4}))
\begin{displaymath}
        j^{\textrm{tor}} = 
        \rho_{e}\,\frac{\Omega_{F}-\omega}{\alpha}\,\tilde{\omega} + 
        \frac{1}{\alpha}\,\frac{dI}{d\Psi}\,\frac{2\,I}{\alpha\,\tilde{\omega}}.
\end{displaymath}
Here, we also eliminate $\rho_{e}$: From (\ref{strau:9.6}) and (\ref{strau:9.2}) 
we deduce
\begin{equation}
        8\pi^{2}\rho = -\mathbf{d}\boldsymbol{*}\left(
         \frac{\Omega_{F}-\omega}{\alpha}\,\mathbf{d}\Psi \right),
        \label{strau:9.12}
\end{equation}
i.e.,
\begin{equation}
        8\pi^{2}\rho_{e} = -\vec{\nabla}\cdot \left(\frac{\Omega_{F}-\omega}{\alpha}\,
        \vec{\nabla}\Psi \right).
        \label{strau:9.13}
\end{equation}
We thus arrive at
\begin{equation}
        j^{\textrm{tor}} = -\frac{1}{8\pi^{2}}\,\frac{\Omega_{F}-\omega}{\alpha}\,
        \tilde{\omega}\,\vec{\nabla}\cdot \left(\frac{\Omega_{F}-\omega}{\alpha}\,
        \vec{\nabla}\Psi \right) + \frac{2\,I}{\alpha^{2}\,\tilde{\omega}}\,
        \frac{dI}{d\Psi}.
        \label{strau:9.14}
\end{equation}

On the other hand, this quantity can also be obtained from Maxwell's 
equation (\ref{strau:9.2}):
\begin{equation}
        4\pi\alpha\,\mathcal{J}^{\textrm{tor}} = [\mathbf{d}(\alpha 
        \boldsymbol{*}\mathcal{B})]^{\textrm{tor}} + [L_{\beta}\boldsymbol{*}
        \mathcal{E}]^{\textrm{tor}}.
        \label{strau:9.15}
\end{equation}
For the first term on the right we have
\begin{equation}
        2\pi\,[\mathbf{d}(\alpha 
        \boldsymbol{*}\mathcal{B})]^{\textrm{tor}} = 2\,\pi\,\mathbf{d}(\alpha 
        \boldsymbol{*}\mathcal{B}^{\textrm{tor}}) \stackrel{(\ref{strau:9.4})}{=}
        \mathbf{d}\,[\alpha \boldsymbol{*}(\mathbf{d}\Psi \wedge \mathbf{d}\varphi)].
        \label{strau:9.16}
\end{equation}

Now, we use the following useful general identity, whose proof is 
left as an exercise: Let $\chi$ be a p-form and $\vec{m}$ a Killing 
field. Assume also $L_{\vec{m}}\,\chi =0$, then
\begin{equation}
        \delta\,(m \wedge \chi) = -m \wedge \delta\,\chi,
        \label{strau:9.17}
\end{equation}
where $\delta$ is the codifferential.

For $\chi=(\alpha/\tilde{\omega}^{2})\,\mathbf{d}\Psi$, 
$\vec{m}=\vec{\partial}_{\varphi}$, $m=\tilde{\omega}^{2}\mathbf{d}\varphi$ 
this gives
\begin{displaymath}
        \delta\,(\alpha\,\mathbf{d}\Psi \wedge \mathbf{d}\varphi) =\delta\, \left( \frac{\alpha}
        {\tilde{\omega}^{2}} \,\mathbf{d}\Psi \right)\,\tilde{\omega}^{2}\,\mathbf{d}\varphi
\end{displaymath}
or
\begin{displaymath}
        \mathbf{d}\boldsymbol{*}[\alpha\,\mathbf{d}\Psi \wedge \mathbf{d}\varphi] =\delta \left( 
        \frac{\alpha}{\tilde{\omega}^{2}} \,\mathbf{d}\Psi \right)\tilde{\omega}^{2}
        \boldsymbol{*}\mathbf{d}\varphi.
\end{displaymath}
Thus, (\ref{strau:9.16}) becomes
\begin{equation}
        2\pi\,\mathbf{d}(\alpha 
        \boldsymbol{*}\mathcal{B}^{\textrm{pol}}) = -\vec{\nabla} \cdot
        \left(  \frac{\alpha}{\tilde{\omega}^{2}} \,\vec{\nabla}\Psi \right)\,
        \tilde{\omega}^{2}      \boldsymbol{*}\mathbf{d}\varphi.
        \label{strau:9.18}
\end{equation}

Next, we turn to the second term in (\ref{strau:9.15}). By (\ref{strau:9.6}) we 
have
\begin{displaymath}
        -L_{\beta}\boldsymbol{*}\mathcal{E} = L_{\beta}\,\left[\frac{\Omega_{F}-\omega}
        {2\pi \alpha}\boldsymbol{*}\mathbf{d}\Psi \right].
\end{displaymath}
If $\chi$ denotes the square bracket, we can write $L_{\beta}\,\chi = 
L_{-\omega\,\partial_{\varphi}}\,\chi = 
-\omega\,L_{\partial_{\varphi}}\,\chi - \mathbf{d}\omega \wedge 
i_{\partial_{\varphi}}\,\chi$, and obtain
\begin{equation}
        \boldsymbol{*}L_{\beta}\boldsymbol{*}\mathcal{E} = - \frac{\Omega_{F}-\omega}
        {2\,\pi\,\alpha}\,i_{\vec{\nabla}\,\omega}\,(\mathbf{d}\Psi \wedge 
        \tilde{\omega}^{2}\,\mathbf{d}\varphi).
        \label{strau:9.19}
\end{equation}
With (\ref{strau:9.18}) and ({\ref{strau:9.19}) we obtain for the Hodge-dual of 
(\ref{strau:9.15})
\begin{displaymath}
        4\pi\alpha\boldsymbol{*}\mathcal{J}^{\textrm{tor}} = 
        -\frac{1}{2\,\pi}\,\vec{\nabla} \cdot
        \left(  \frac{\alpha}{\tilde{\omega}^{2}} \,\vec{\nabla}\Psi \right)\,
        \tilde{\omega}^{2}\,\mathbf{d}\varphi - \frac{\Omega_{F}-\omega}
        {2\,\pi\,\alpha}\,\left(\vec{\nabla}\,\omega \cdot \vec{\nabla}\,\Psi
        \right)\, 
        \tilde{\omega}^{2}\,\mathbf{d}\varphi.
\end{displaymath}
Since $\boldsymbol{*}\mathcal{J}^{\textrm{tor}} = 
j^{\textrm{tor}}\,\frac{1}{\tilde{\omega}}\,\tilde{\omega}^{2}\,\mathbf{d}\varphi$, 
we get
\begin{equation}
        \frac{4\pi\alpha}{\tilde{\omega}}\,j^{\textrm{tor}} = 
        -\frac{1}{2\,\pi}\,\vec{\nabla} \cdot
        \left(  \frac{\alpha}{\tilde{\omega}^{2}} \,\vec{\nabla}\Psi \right) -
        \frac{\Omega_{F}-\omega}
        {2\,\pi\,\alpha}\,\left(\vec{\nabla}\,\omega \cdot \vec{\nabla}\,\Psi
        \right).
        \label{strau:9.20}
\end{equation}
Inserting $\vec{\nabla}\omega = \vec{\nabla}(\omega-\Omega_{F}) + 
(d\Omega_{F}/d\Psi)\,\vec{\nabla}\Psi$ leads finally to our second 
formula for $j^{\textrm{tor}}$:
\begin{eqnarray}
        8\pi^{2}j^{\textrm{tor}} & = & -\frac{\tilde{\omega}}{\alpha}\,\vec{\nabla} \cdot
        \left(  \frac{\alpha}{\tilde{\omega}^{2}} \,\vec{\nabla}\Psi 
        \right) +       
        \frac{\tilde{\omega}}{\alpha^{2}}\,(\Omega_{F}-\omega)\,\vec{\nabla}\Psi 
        \cdot \vec{\nabla}(\Omega_{F}-\omega)
        \nonumber  \\
         &  &  
         -\frac{\tilde{\omega}}{\alpha^{2}}\,(\Omega_{F}-\omega)\,\frac{d\Omega_{F}}
         {d\Psi}
         \,(\vec{\nabla}\Psi)^{2}.
        \label{strau:9.21}
\end{eqnarray}
Comparison of (\ref{strau:9.15}) with (\ref{strau:9.21}) gives, after a few steps, 
the following \emph{generalized Grad-Shafranov equation}:
\begin{equation}
        \vec{\nabla} \cdot \left\{ \frac{\alpha}{\tilde{\omega}^{2}}\left[ 
        1 - 
        \frac{(\Omega_{F}-\omega)^{2}\,\tilde{\omega}^{2}}{\alpha^{2}}\right] \vec{\nabla}\Psi
        \right\} + \frac{\Omega_{F}-\omega}{\alpha}\, \frac{d\Omega_{F}}{d\Psi}
         \,(\vec{\nabla}\Psi)^{2} + 
         \frac{16\,\pi^{2}}{\alpha\,\tilde{\omega}^{2}} 
         \,I\,\frac{dI}{d\Psi} = 0.
        \label{strau:9.22}
\end{equation}

The integration of the original equations (for axisymmetric stationary 
situations) is reduced to this single partial differential equation. 
$I(\Psi)$ and $\Omega_{F}(\Psi)$ are free functions\footnote{These 
are, of course, restricted by boundary conditions, but we do not 
discuss this here.}, and if $\Psi$ is a solution of (\ref{strau:9.22}) the 
electromagnetic fields are given by (\ref{strau:9.4}) and (\ref{strau:9.6}), while 
the charge and current distribution can be obtained from 
(\ref{strau:9.13}), (\ref{strau:9.7}), and (\ref{strau:9.14}).

A limiting case of (\ref{strau:9.22}) is known from the electrodynamics of 
pulsars: Ignoring the curvature of spacetime and using that 
$\Omega_{F}$ is equal to the angular velocity $\Omega$ of the neutron 
star (to derived shortly), we get the \emph{pulsar equation}:
\begin{equation}
        \vec{\nabla} \cdot \left\{ \frac{1}{\tilde{\omega}^{2}}\left[ 
        1 - \Omega^{2}\,\tilde{\omega}^{2}\right] \vec{\nabla}\Psi
        \right\} +       \frac{16\,\pi^{2}}{\tilde{\omega}^{2}} 
         \,I\,\frac{dI}{d\Psi} = 0
        \label{strau:9.23}
\end{equation}
($\tilde{\omega}$ is the radial cylindrical coordinate in flat space).

The equation (\ref{strau:9.22}) holds also outside an aligned pulsar, since 
all the basic equations in section 2 remain valid there. However, the 
different boundary condition at the surface of the neutron star 
implies $\Omega_{F}=\Omega$, as we now show. In the interior of the 
neutron star the 3-velocity is (see (\ref{strau:8.11}))
\begin{equation}
        \vec{v} = 
        \frac{1}{\alpha}\,(\Omega-\omega)\,\vec{\partial}_{\varphi} .
        \label{strau:9.24}
\end{equation}
Since the neutron star matter is ideally conducting, the electric 
field there is
\begin{displaymath}
        \mathcal{E} = i_{\vec{v}}\,\mathcal{B} = 
        -\frac{1}{2\pi\alpha}\,(\Omega-\omega)\, \mathbf{d}\Psi,
\end{displaymath}
if the $\mathcal{B}$-field is poloidal (first term in (\ref{strau:9.4})). At 
the boundary this has to agree with (\ref{strau:9.6}), implying $\Omega_{F}=\Omega$.

Another remark should be made at this point about the interior of the 
neutron star. We showed in section 8 that $\mathbf{d}\Omega_{F} \wedge \mathbf{d}\Psi = 
0$ implying $\mathbf{d}\Omega \wedge \mathbf{d}\Psi = 0$ inside the star. It will, in 
general, not be possible to represent $\Psi$ as a function of 
$\Omega$, since $\Psi$ has to satisfy the Grad-Shafranov equation for 
$I(\Psi) =0$ inside the star. Therefore, the rotation must be 
\emph{rigid}, $\Omega = \textrm{const}$.

\end{document}